\documentclass[fleqn,usenatbib]{mnras}

\usepackage{newtxtext,newtxmath}
\usepackage[T1]{fontenc}

\DeclareRobustCommand{\VAN}[3]{#2}
\let\VANthebibliography\thebibliography
\def\thebibliography{\DeclareRobustCommand{\VAN}[3]{##3}\VANthebibliography}

\usepackage{graphicx}	% Including figure files
\usepackage{amsmath}	% Advanced maths commands

\title[Infall driven gravitational instability]{Infall-driven gravitational instability in accretion discs}% --- I. Self regulation and spiral mode analysis} 
\author[Longarini et al.]{
Cristiano Longarini,$^{1}$\thanks{email:\url{ cl2000@cam.ac.uk}} 
Daniel J. Price,$^{2}$
Kaitlin M. Kratter,$^{3}$
Giuseppe Lodato$^{4}$ and
Cathie J. Clarke$^{1}$
\\
$^{1}$Institute of Astronomy, University of Cambridge, Madingley Road, Cambridge, CB3 0HA, United Kingdom\\
$^{2}$School of Physics and Astronomy, Monash University, Clayton, VIC 3800, Australia\\
$^{3}$Department of Astronomy and Steward Observatory, University of Arizona, 933 N Cherry Ave, Tucson, AZ, 85721, USA\\
$^{4}$Dipartimento di Fisica, Università degli Studi di Milano, Via Celoria 16, 20133 Milano, Italy
}

\date{Accepted XXX. Received YYY; in original form ZZZ}

\pubyear{\the\year{}}

\begin{document}
\label{firstpage}
\pagerange{\pageref{firstpage}--\pageref{lastpage}}
\maketitle

% Abstract of the paper
\begin{abstract}
Gravitational instability (GI) is typically studied in cooling-dominated discs, often modelled using simplified prescriptions such as $\beta$-cooling. In this paper, we investigate the onset and evolution of GI in accretion discs subject to continuous mass injection, combining 1D and 3D numerical simulations. We explore an alternative self-regulation mechanism in which mass replenishment drives the system toward marginal stability $Q\sim 1$. In this regime, the disc establishes a steady-state disc-to-star mass ratio, balancing the mass transported to the central object with that added to the disc. Our 3D simulations reveal that the general scaling predicted from the linear theory are respected, however there are important difference compared to the cooling case in terms of morphology and pattern speed. Unlike the flocculent spirals seen in cooling-driven instability, the power is concentrated towards the dominant modes in infall-driven spirals. Additionally, spiral waves generate at the mass injection location, and propagate at constant pattern speed, unlike in the cooling case. This suggests a fundamental difference in how mass-regulated and cooling-regulated discs behave and transport angular momentum.
\end{abstract}

% Select between one and six entries from the list of approved keywords.
% Don't make up new ones.

%%%%%%%%%%%%%%%%%%%%%%%%%%%%%%%%%%%%%%%%%%%%%%%%%%

%%%%%%%%%%%%%%%%% BODY OF PAPER %%%%%%%%%%%%%%%%%%

\section{Introduction}
Young protoplanetary discs are likely to be massive, as most of their mass has not yet been accreted by the central object. Observations support this, as seen in the cumulative mass function of protostellar discs across different star-forming regions and ages \citep{tobin20}. Young and massive discs are likely to be gravitationally unstable \citep{kratter16}. The instability threshold for axisymmetric perturbations is determined by the dimensionless Toomre parameter \citep{safronov60,toomre64,lin64}, according to
\begin{equation}\label{toomreQ}
    Q = \frac{c_s \kappa}{\pi G \Sigma},
\end{equation}
where $c_s$ is the sound speed, $\Sigma$ the disc surface density and $\kappa$ the epicyclic frequency, with $Q=1$ the instability threshold.

In numerical simulations, gravitational instability is usually triggered by cooling the disc. The most widely used cooling prescription is the simple \(\beta\)-cooling model, in which the disc cools on a timescale proportional to the local dynamical time, with a proportionality factor \(\beta\) \citep{gammie01}. For cooling-driven gravitational instability, the system can reach a self-regulated thermal state. An initially hot and stable disc (\(Q > 1\)) cools according to the \(\beta\)-cooling prescription until it reaches the instability threshold (Q = 1). At this point, gravitational instability sets in, triggering the formation of spiral waves. These waves heat the disc through shocks and increase the \(Q\)-parameter. Eventually, the heating from the spiral arms balances the imposed cooling, leading to a self-regulated state. In this scenario, gravitational instability acts as a \emph{thermostat}, maintaining the disc around \(Q = 1\).

The $\beta-$cooling model has proved an effective sandbox for studying self-gravitating discs, since one is able to study gravitationally unstable discs in a quasi-steady, equilibrium state. This has enabled detailed study of the angular momentum transport \citep{cossins09,lodato07}, dust accumulation \citep{rice04,rice06,booth16,baehr21,longarini23b,rowther24a}, kinematic perturbations \citep{hall20,longarini21,terry22}. The main problem is that the $\beta-$cooling prescription does not account for heating by the central star. A number of studies have attempted to account for irradiation in discs \citep{boss02,mejia04,boley07,stamatellos07,forgan09,young24}. In particular, \cite{rice11} and more recently \cite{leedham25} have shown that thermal self-regulation can be ineffective in strongly irradiated discs, whose behaviour is similar to polytropic models originally studied by \cite{laughlin96}. Such studies have been confirmed in the global simulations with the live radiative transfer of \cite{rowther24b}, who finds negligible shock heating in their irradiated discs.

In the context of gravitational instability, potential candidates include Elias 2-27 \citep{perez16,longarini24}, AB Aurigae \citep{fukugawa04,tang12}, and GM Aur \citep{martire24}. Interestingly, these sources show signs of infall \citep{paneque21,speedie24,schwarz21}. Adding mass to a protoplanetary disc can be an alternative way to trigger gravitational instability \citep{kratter08,kratter10a}. Considerable theoretical effort has been devoted to understanding the dynamics and consequences of infall on protoplanetary discs \citep[e.g.][]{cameron78,linpringle90,kenyon93,bate03,kratter08,zhu10,winter24,calcino25}. Observations indicate that this process remains significant even in evolved stages of protoplanetary discs lifetime, suggesting the relevance of late infall \citep{gupta23,gupta24,winter24b}. Adding mass from the environment offers an alternative way for a disc to self-regulate. In the absence of other instabilities and with a fixed temperature profile, a disc will passively accept mass until the $Q=1$ threshold is reached. Then, spiral arms will form and angular momentum transport will start, leading to a balance between the mass infalling from the environment, and mass being drained onto the star, decreasing $\Sigma$ and hence increasing $Q$.% A \emph{bathtub} rather than a thermostat. 

Our aim in this paper is to explore whether a steady state can be reached in discs that are not isolated from their environment. In particular, we use a combination of 1D and 3D simulations to examine how the long term behaviour of infall-regulated gravitationally unstable discs differs from their temperature-regulated counterparts. %We show that the resultant spiral structure is similar to that predicted from $\beta-$cooling simulations. This explains the success of the $\beta$-cooling approach, while offering a more realistic method of comparing simulations to observations.

The paper is organised as follows: in Section~\ref{section2} we present an analytical framework to understand how self-regulation operates in discs undergoing infall; Section~\ref{section3} describes the numerical framework including 1D and 3D simulations. Section~\ref{section4} presents the numerical results, and a comparison between the two approaches. In Section~\ref{section5} we discuss possible observational signatures and possible correlations in the initials condition for planet formation. While our focus is on protostellar discs, the methods, scalings and conclusions are general and can be applied to self-gravitating accretion discs at any scale, such as those around supermassive black holes. 

\section{Gravitational instability and the role of infall}\label{section2}
The linear evolution of gravitational instability is described by the dispersion relation \citep{safronov60,lin64,toomre64}  
\begin{equation}\label{disprel}
    (\omega - m\Omega)^2 = c_s^2 k^2 - 2\pi G \Sigma |k| + \kappa^2,
\end{equation}  
where \(\omega\) is the perturbation frequency, \(m\) is the azimuthal wavenumber, and \(\kappa\) is the epicyclic frequency, defined as 
\begin{equation}\label{epicyclic}
    \kappa^2 = \frac{2\Omega}{R}\frac{\rm d}{\text{d}R}\left(\Omega R^2\right),
\end{equation}
being equal to the orbital frequency $\Omega$ for a Keplerian disc. This relation is valid in the WKB approximation, which requires the radial wavelength of the perturbation to be much smaller than the azimuthal one, i.e., \(|kR| \gg m\). The stability of axisymmetric perturbation is controlled by the Toomre parameter, as defined in Eq.~\eqref{toomreQ}, with instability occurring when \(Q \leq 1\). In an unstable disc, the most unstable mode corresponds to a radial wavenumber  
\begin{equation}\label{kmax}
    k_{\rm max} = \frac{\pi G \Sigma}{c_s^2}.
\end{equation}  
For a marginally unstable disc $Q=1$,  a spiral perturbation with a radial wavenumber $k=k_\text{max}$ satisfies the dispersion relation in Eq.~\eqref{disprel} with the left-hand side equal to zero. This means that the perturbation satisfies  $\omega = m\Omega$, leading to a pattern speed  
\begin{equation}
    \Omega_p = \frac{\omega}{m} = \Omega.
\end{equation}  
Under these conditions, we expect spiral density waves to generate at co-rotation, and have an instantaneous pattern speed of the order of the Keplerian frequency. 

In general, for a given pattern speed $\Omega_p$, it is possible to define the Lindblad resonances, that are the location where the gravitational perturbation of the spiral matches the natural epicyclic frequency of the disc material, and the corotation resonance, where the pattern speed matches the angular frequency of the disc. In the case of a constant pattern speed $\Omega_p$ in a Keplerian disc, the radii of the inner and outer Lindblad resonances are \citep[e.g.][]{ogilvie02})
\begin{equation}\label{lindblad_resonances_keplerian}
    R_{\rm ILR,OLR} = R_{\rm cr}\left(1\pm \frac{1}{m}\right)^{2/3}.
\end{equation}
 
In non-axisymmetric self-gravitating discs, torques induced by gravitational perturbations transport both energy and angular momentum. Understanding how this happens requires studying the non-linear evolution of gravitational instability. The ability of self-gravitating spirals to transport angular momentum was first identified by \cite{lyndenbell72} in the context of galactic dynamics. In protostellar discs, \cite{balbus99} computed the stress tensor associated with gravitational potential perturbations, noting that energy transport via gravitational instability cannot be fully described by a local viscous approximation—except at co-rotation. This can be understood considering the spiral wave energy per unit surface $\mathcal{E}_{\rm w}$, which reads
\begin{equation}\label{wave_dens}
    \mathcal{E}_{\rm w} = \frac{\Sigma}{2}\frac{m^2}{k^2}(\Omega-\Omega_p)^2\left(\frac{\delta \Sigma}{\Sigma}\right)^2 + \frac{\Sigma}{2}\frac{m^2}{k^2}\Omega(\Omega-\Omega_p)\left(\frac{\delta \Sigma}{\Sigma}\right)^2.
\end{equation}
The second term --- given by the angular momentum per unit area multiplied by the rotation speed --- represents a local energy transport process, and can therefore be modelled using the standard $\alpha$ prescription. In contrast, the first term --- equal to the same angular momentum term times $\Omega-\Omega_p$ --- is a non-local term, and prevents GI discs from behaving as pure $\alpha-$discs. 
The degree of non-local angular momentum transport %can be quantified as the ratio between the global and the local term in Eq.~\eqref{wave_dens} \citep{cossins09}
%\begin{equation}\label{nonlocal_eq}
 %   \xi = \left|\frac{\Omega-\Omega_p}{\Omega}\right|, 
%\end{equation}
%where $\xi$ corresponds to the distance a perturbation can travel far from co-rotation, ultimately determining the extent of non-locality of angular momentum transport. The quantity that ultimately determines it is
is ultimately determined by the disc-to-star mass ratio (e.g. \cite{laughlin94, laughlin96,laughlin97,laughlin98}), with more massive discs globally redistributing angular momentum within the disc.

\subsection{Mass controlled self-regulation}

An initially stable disc $Q\gg 1$ does not show any spiral density wave. To trigger gravitational instability, we can either cool the disc, i.e. decrease its sound speed, or add mass, i.e. increase the surface density. In the case of infall-driven GI, the instability is triggered by mass injection, and self-regulation is obtained in terms of mass, rather than temperature. The self-regulated state is characterised by a stationary $Q\approx1$ profile, meaning that the time derivative of the Toomre parameter is equal to zero. This implies that the injected mass is distributed between the central object and the disc, to ensure that $\dot{Q}=0$. It is possible to obtain the balance mentioned above as follows: denoting $\dot{M}_{\rm inj}$ as the mass injection rate, $\dot{M}_{\star}$ as the mass accretion rate onto the central object and $\dot{M}_{d}$ {the rate at which the disc mass is increasing}, mass conservation implies that
\begin{equation}\label{mass_cons}
    \dot{M}_{\rm inj} = \dot{M}_{\star} + \dot{M}_{d}.
\end{equation}
As mentioned above, the evolution is such that the time derivative of the Toomre parameter is zero. The Toomre parameter can be re-written as 
\begin{equation}
    Q = A\frac{H}{R}\frac{M_\star}{M_d},
\end{equation}
where $A$ is a constant of order unity and $H \equiv c_{\rm s}/\Omega$ is the pressure scale height. As a working hypothesis, we suppose that the disc is locally isothermal, hence the temperature does not change with time, and we also suppose that the disc scale height is constant $H/R\sim \rm const$. %\footnote{In principle, $H/R\propto M_\star^{-1/2}$. However, we don't expect the star mass to evolve so fast to make a difference in this respect. In additon, in the 1D code used for this work, we take into account this scaling, and we find that the results are not impacted by the hypothesis $H/R\sim \rm const$. }. 
Within these hypotheses, the steady state condition is
\begin{equation}\label{toomre_stat}
    \dot{Q} = 0 \to \frac{\dot{M}_d}{M_d} = \frac{\dot{M}_\star}{M_\star} .
\end{equation}
Combining Eqs.~\eqref{mass_cons} and \eqref{toomre_stat}, the evolution of the stellar and disc mass is given by
\begin{equation}\label{mdot_star_SG}
    \dot{M}_{\star} = \frac{\dot{M}_{\rm inj}}{1+q},
\end{equation}
and
\begin{equation}\label{mdot_disc_SG}
    \dot{M}_{\rm d} = \dot{M}_{\rm inj}\frac{q}{1+q},
\end{equation}
where $q=M_d/M_\star$. These are the key characteristics of self-regulation: mass is injected into the disc at a specific location and at a given rate, and it is redistributed between the star and the disc to keep their ratio constant over time. To achieve this, Eqs.~\eqref{mdot_star_SG} and \ref{mdot_disc_SG} must be respected. 

Since the problem is completely rescalable in terms of star mass, disc mass and injection rate, it is useful to define a dimensionless injection rate, which does not depend on the physical scale of the system. The dimensionless injection rate $\dot{\mu}$ is defined as
\begin{equation}\label{dotmu}
    \dot{\mu}=\frac{\dot{M}_\text{inj}}{M_\star\Omega_\text{inj}}.
\end{equation}

%This process can be  described using the analogy of a faucet and a sink, rather than a thermostat as in the cooling scenario. The injected mass acts like water flowing from a faucet into a sink, while the central star behaves as a drain, accreting the mass. When the mass injection increases (the water flow from the faucet intensifies), the instability grows, resulting in high accretion rates (fast drainage). However, if the injection becomes excessive, the system can fragment, analogous to the sink overflowing. Conversely, when the mass injection decreases, the accretion rate slows down. Self-regulation in this scenario is akin to a well-calibrated faucet and drain system that prevents the sink from either overflowing or emptying out completely. Therefore, the equivalent self-regulated condition corresponds to the mass injection rate matching the star's accretion rate.

where $\Omega_{\rm inj} = \Omega(R_{\rm inj})$, being $R_{\rm inj}$ being the injection radius. According to the stationary solution for a self-gravitating accretion disc \citep{bertin97,bertin99,lodato07}, the mass accretion rate is related to the $\alpha$, i.e. the efficiency of angular momentum transported throughout the disc, according to 
\begin{equation}
     \dot{M}_\star = \frac{2\alpha c_s^3}{G}\left| \frac{\rm d \log \Omega}{\rm d \log R} \right| = \frac{3\alpha c_s^3}{GQ},
\end{equation}
where in the second equality we assumed that the disc is Keplerian. Hence, using the self-regulation condition of Eq.~\eqref{mdot_star_SG} the $\alpha$ viscosity can be written as
\begin{equation}\label{alpha_TH}
    \alpha =  \frac{\dot{M}_{\rm inj}}{M_\star \Omega}\frac{1}{3(H/R)^3(1+q)} = \frac{\dot{\mu}}{3(H/R)^3(1+q)}.
\end{equation}

\section{Numerical methods}\label{section3}
In this section we present the numerical framework of this paper. We describe the 1D and the 3D codes, together with the set of simulations we performed. 

\subsection{1D grid evolution code}
We evolve the surface density of an accretion disc undergoing mass injection with a 1D evolution code\footnote{\url{https://github.com/crislong/discfusion}}. The diffusion equation for the surface density of a Keplerian disc with mass injection is \citep{lyndenbell72,linpringle90}
\begin{equation}\label{diff_eq_dimensional}
    \frac{\partial \Sigma}{\partial t} = \frac{3}{R}\frac{\partial}{\partial R}\left[\sqrt{R}\frac{\partial}{\partial R}\left(\sqrt{R}\Sigma \nu\right)   \right] + \dot{\Sigma}_{\rm inj},
\end{equation}
where $\dot{\Sigma}_{\rm inj}$ is a source term describing the mass injection and $\nu$ is the kinematic viscosity. For simplicity, we suppose that mass is added at a given radial location in the disc $R_{\rm inj}$ with a constant mass injection rate $\dot{M}_{\rm inj}$, according to 
\begin{equation}
    \dot{\Sigma}_{\rm inj} = \frac{\dot{M}_{\rm inj}}{2\pi R_{\rm inj}} \delta(R-R_{\rm inj}).
\end{equation}
To solve Eq.~\eqref{diff_eq_dimensional} numerically we follow \cite{bath81}. Details can be found in Appendix~\ref{sec:1Dcode}. To mimic the effect of gravitational instability, we force the kinematic viscosity to switch on only when the Toomre parameter $Q$ is below a given threshold $Q_c$, according to \citep{linpringle87,linpringle90}
\begin{equation}
    v_{\rm GI} =
\begin{cases} 
\eta \left( \frac{Q_c^2}{Q^2} - 1 \right) \left( \frac{c_s^2}{\Omega} \right), & Q \leq Q_c \\ 
0, & \text{otherwise}.
\end{cases}
\end{equation}
We choose $Q_c=1$ and $\eta=0.5$; we tested different values of this parameter $\eta = 0.1, 1$, and we found that the results do not change, as found by \cite{lodatonat06}, who used the same picture of infall dominated accretion to explain the formation of supermassive black hole seeds in the early Universe.

Since in the simulation the star mass evolves because of accretion, at each time-step we update its value and as a consequence $\Omega, Q$ and $\nu$. 

We tested different viscosity prescription for gravitationally unstable discs \citep{kratter08,rafikov15}. The results are presented in Appendix~\ref{app_viscosity}. The take home message is that the overall evolution is not impacted when using more refined viscosity models.

\subsubsection{Numerical setup}
\begin{table}\centering
\caption{Parameters of the 1D simulations: injection rate  $\dot{M}_{\rm inj}$, dimensionless injection rate, injection location $\hat{R}_{\rm inj}$ and aspect ratio $H/R$.}
{\begin{tabular}{lllll}
\textbf{Simulation} & $\dot{\text{M}}_{\rm inj}$ [M$_\odot$/yr] & $\dot{\mu}$ & $\hat{R}_{\rm inj}$ & $H/R$  \\ \hline
\textbf{S1D\_1} - reference & $5\times10^{-5}$& $2.5\times10^{-4}$  & $1$   & $0.1$ \\
%\textbf{L1} & $10^{-5}$  & $1$   & $0.05$ & LISO \\
\textbf{S1D\_2} & $5\times10^{-5}$ & $2.5\times10^{-4}$  & $1$   & $0.05$ \\
%\textbf{L2} & $10^{-5}$  & $1$   & $0.01$ & LISO \\
\textbf{S1D\_3} & $5\times10^{-5}$& $2.5\times10^{-4}$  & $1$   & $0.15$  \\
%\textbf{L3} & $10^{-5}$  & $1$   & $0.1$  & LISO \\
\textbf{S1D\_4} & $10^{-6}$& $5\times10^{-6}$  & $1$ & $0.1$ \\
%\textbf{L4} & $10^{-5}$  & $0.2$ & $0.05$ & LISO \\
\textbf{S1D\_5} & $10^{-5}$& $5\times10^{-5}$  & $1$ & $0.1$ \\
%\textbf{L5} & $10^{-5}$  & $0.5$ & $0.05$ & LISO \\
\textbf{S1D\_6} & $10^{-4}$& $5\times10^{-4}$  & $1$   & $0.1$ \\
%\textbf{L6} & $10^{-6}$  & $1$   & $0.05$ & LISO \\
\textbf{S1D\_7} & $5 \times 10^{-5}$& $2.5\times10^{-4}$ & $0.5$ & $0.1$ \\
%\textbf{L7} & $5 \times 10^{-6}$ & $1$ & $0.05$ & LISO \\
\textbf{S1D\_8} & $5 \times 10^{-5}$& $2.5\times10^{-4}$ & $0.2$ & $0.1$ \\
%\textbf{L8} & $5 \times 10^{-5}$ & $1$ & $0.05$ & LISO \\
%\textbf{S1D\_9} & $5 \times 10^{-5}$  & $0.2$   & $0.1$ \\
%\textbf{L9} & $10^{-4}$  & $1$   & $0.05$ & LISO \\
\end{tabular}}\label{simulations_table_1D}
\end{table}

We perform simulations assuming a locally isothermal disc with the sound speed $c_s$ constant with time and with radius. We explore the parameters $\dot{M}_{\rm inj}$, $H/R$ and $\hat{R}_{\rm inj} = R_{\rm inj}/R_{\rm out}$. Table~\ref{simulations_table_1D} lists the values used in each simulation. We initialise a disc with a surface density profile $\Sigma \propto R^{-1.5}$, initially extending from $R_{\rm in} = 1$ to $R_{\rm out}=10$ in code units. The linear radial grid extends from 1 to $10^3$ in code units, to allow the disc to expand without reaching the outer boundary. The resolution is set by the number of radial zones, that is $2\times10^3$. The initial disc mass is chosen so that the Toomre parameter at the injection radius is initially $1.1$. The inner and outer boundary conditions are set so that $\Sigma(1) = \Sigma(10^3) = 0$. We evolved the simulations for 5 injection times, where we define the injection time as the star mass doubling time
\begin{equation}
    t_{\rm inj} \equiv \frac{M_{\star,0}}{\dot{M}_{\rm inj}}. \label{eq:tinj}
\end{equation}
Table~\ref{simulations_table_1D} describes the set of simulations we performed, where \textbf{S1D\_1} is the reference simulation. Starting from its parameters, we vary the aspect ratio, injection rate and injection location. 

In the following plots we will show the evolution up to 2 injection times, since the results afterwards do not change. 

\subsection{3D SPH simulations}
We perform a suite of 3D smoothed particle hydrodynamics (SPH; \citealt{lucy77,gingold77,price12}) simulations of gas discs using the code \textsc{Phantom} \citep{price18}. The self-gravity in {\sc Phantom} is described in the code paper, with the algorithm for softening the short range interaction described in \citet{pricemonaghan07}. This code has been widely used in the astrophysics community to study gas dynamics in accretion discs \citep[e.g.][]{lodato10,cuello19,ragusa20,nealon22,ceppi23}, and recently has been used for studying gravitationally unstable discs \citep[e.g.][]{hall20, cadman22,rowther22,longarini23b}. To date, gravitational instability has been numerically triggered by cooling the disc, using the simple $\beta-$prescription \citep{gammie01}, or variations thereof \citep{stamatellos07,young24}. In this work, we trigger gravitational instability by adding mass with a chosen angular momentum to a disc with a prescribed vertically and locally isothermal temperature profile. While previous numerical simulations have explored infall-driven gravitational instability \citep{kratter10a}, these studies focused on simulating the collapse of a quasi-spherical core mediated by an accretion disc. In contrast, we consider a more controlled experiment with two free parameters by injecting SPH particles directly into the simulation, without modelling the collapse of a gas sphere.
%In this work, we adopt a simpler setup, making no assumptions about the geometry or origin of the infalling mass. Instead, the system is governed by two key parameters that control the dynamics.

We inject SPH particles at the injection radius $R_{\rm inj}$ with a constant mass injection rate $\dot{M}_{\rm inj}$, at Keplerian velocity $v_{\rm inj} = \sqrt{GM_\star/R_{\rm inj}}$ and distributed vertically according to a Gaussian distribution with standard deviation equal to the pressure scale height $H$, which we choose by setting the aspect ratio $(H/R)_{\rm inj}$ to match that of the disc at the injection radius. To minimise asymmetries during the injection process, which could cause a secular displacement of the system's centre of mass, we add particles in pairs with symmetric positions relative to the central object. The injection module is publicly available in {\sc Phantom} v2025.0.0 or higher\footnote{\url{https://github.com/danieljprice/phantom}} using the \verb+isosgdisc+ setup.

\subsubsection{Numerical setup}
\begin{table}\centering\caption{Parameters of the 3D simulations: injection rate, dimensionless injection rate, injection location and aspect ratio.}
{\begin{tabular}{lllll}
\textbf{Simulation} & $\dot{\text{M}}_{\rm inj}$ [M$_\odot$/yr] & $\dot{\mu}$ & $\hat{R}_{\rm inj}$ & $H/R$  \\ \hline
\textbf{S3D\_1} - reference  &  $5\times10^{-5}$& $2.5\times10^{-4}$ &  $1$  & $0.1$   \\
\textbf{S3D\_2}  &  $1.25\times10^{-5}$& $5\times10^{-5}$& $1$  & $0.075$   \\
\textbf{S3D\_3}  &  $1\times10^{-4}$& $5\times10^{-4}$& $1$  & $0.125$   \\
%??\textbf{S3D\_4}  &  $10^{-5}$& $1$  & $0.1$   \\
\\
\textbf{Cool\_0325} & $\beta=10$ & $q=0.325$ & \\

\end{tabular}}\label{table_3D}
\end{table}
3D hydrodynamical simulations with mass injection are computationally expensive, as the resolution increases over time due to the continuous addition of particles, and the typical timescale --- the injection timescale --- is long. To address this, we first verify that 1D simulations accurately reproduce the global quantities ($\dot{M}_\star, \dot{M}_d, \dot{q}$). Consequently, we rely on the 1D code for a comprehensive exploration of the parameter space. To complement this, we conduct three \textsc{Phantom} simulations with varying aspect ratios to perform a Fourier analysis of the spiral modes. Additionally, we run a simulation with cooling rather than infall, for morphological comparison.%Additionally, we carry out a simulation with a different injection rate to investigate the resulting spiral arm density contrast. 
Table~\ref{table_3D} summarises the set of simulations.

We initially set up the disc around a central sink particle with mass $M_\star = 1$ in code units and accretion radius $R_{\rm acc}=1$. The disc extends between $R_{\rm in}=1$ and $R_{\rm out}=10$, with a power law surface density $\Sigma\propto R^{p}$, with $p=-1.5$. We assume that the disc is locally isothermal, with the sound speed constant with time and radius. We set the initial disc mass by choosing the initial value of $Q$ to be $1.1$ at the injection radius. The setup is the same as the 1D simulations, and so it is the reference simulation (see Table~\ref{table_3D}). In the simulations we are not using the disc-viscosity flag \citep{lodato10}, meaning that the viscosity is not described by an $\alpha_{\rm SS}$ coefficient. We did so since in these systems the main driver of angular momentum transport is gravitational instability through spiral arms. We instead apply dissipation only as necessary for shock capturing, with the shock viscosity coefficient $\alpha_{\rm AV}\in [0, 1]$ using the \citet{cullen10} switch as described in \citet{price18} and $\beta_{\rm AV} =2$. The initial number of particles is $N_{\rm in} = 5\times10^5$, and during the simulations the number significantly increases, reaching a final value of active particles $N_{\rm f} \gtrsim 1\times10^6$.

As a comparison with the infall driven GI, we perform an additional simulation where gravitational instability is triggered by cooling \textbf{Cool\_0325}. We choose the same disc structure of the reference simulation, except for the inner radius that is set $R_{\rm in} =0.2$. The disc to star mass ratio of the simulation is chosen to match the self-regulated value of the reference simulation, that is $q=0.325$.

\section{Results}\label{section4}

\begin{figure*}
    \includegraphics[width = 0.8\textwidth]{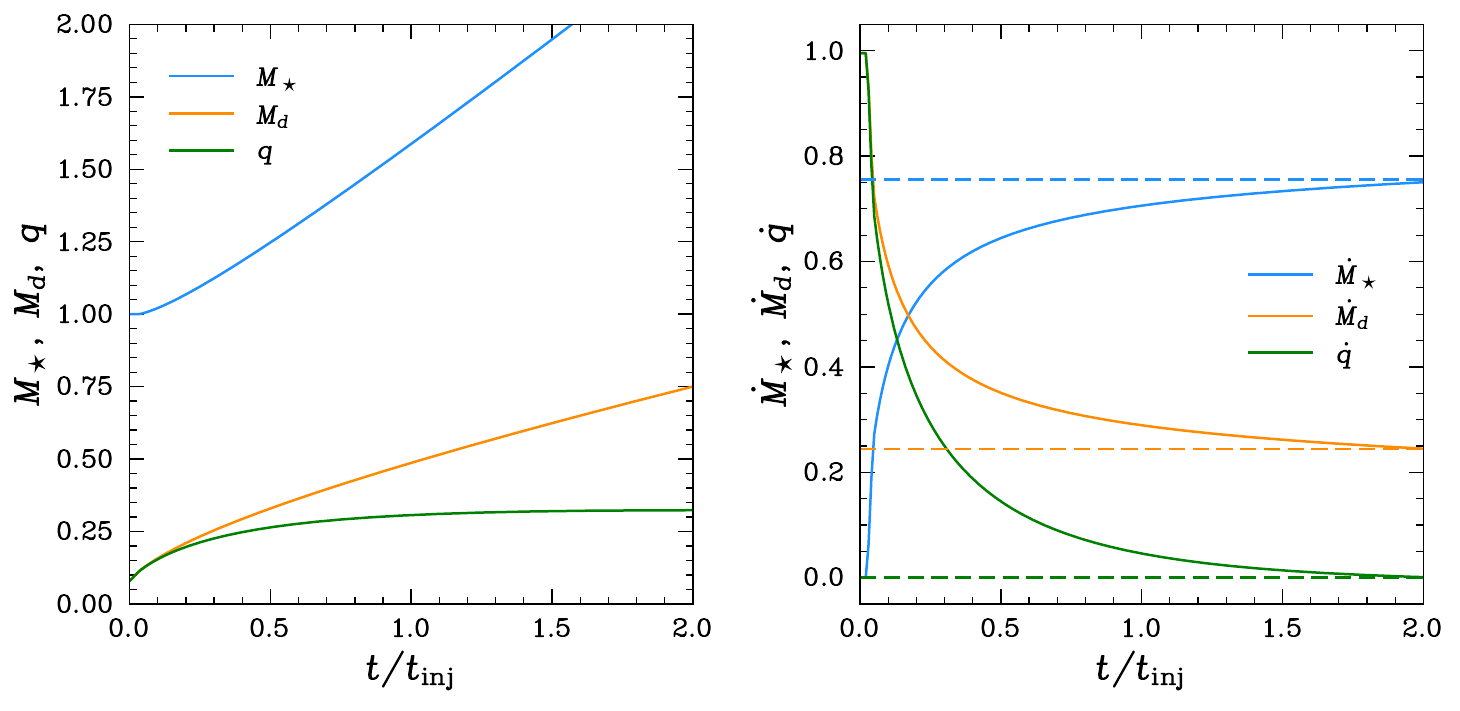}
    \caption{{Left panel}: time evolution of the stellar mass, disc mass and disc to star mass ratio as a function of time in units of injection time. {Right panel}: stellar accretion rate ($\dot{M}_\star$), disc accretion rate ($\dot{M}_{\rm d}$) and time derivative of the disc to star mass ratio ($\dot{q}$) in units of mass injection rate as a function of time in units of injection time. The dashed lines correspond to the steady state prediction: Eq.~ \eqref{mdot_star_SG} for $\dot{M}_\star$, Eq.~\eqref{mdot_disc_SG} for $\dot{M}_{\rm d}$ and zero for $\dot{q}$.}\label{fig:G1_img1}
\end{figure*}

\begin{figure*}
    \includegraphics[width = 0.825\textwidth]{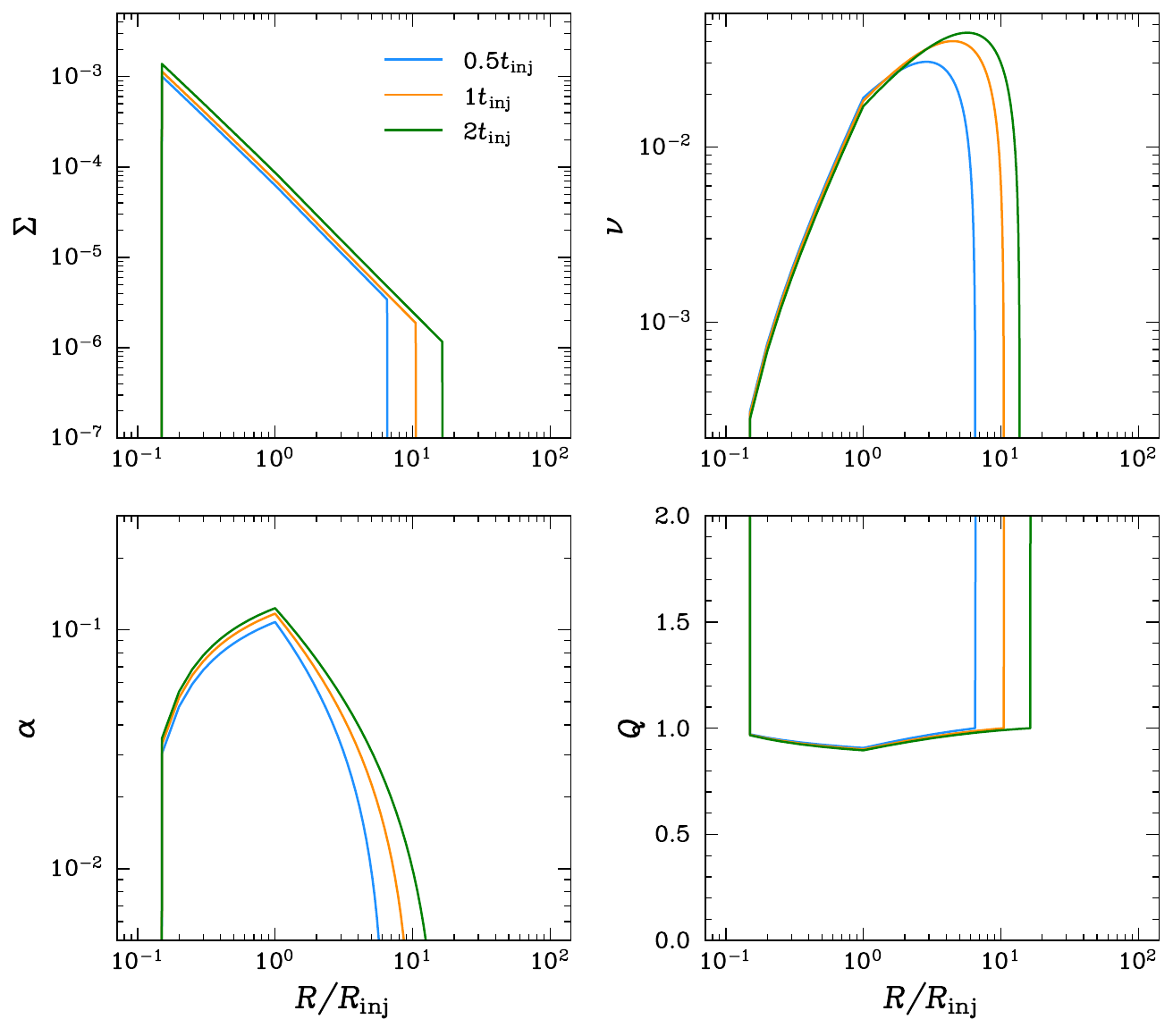}
    \caption{Surface density, kinematic viscosity, alpha viscosity and Toomre Q parameter at different times for the reference simulation.}\label{fig:collection}
\end{figure*}

\begin{figure*}
    \includegraphics[width = \textwidth]{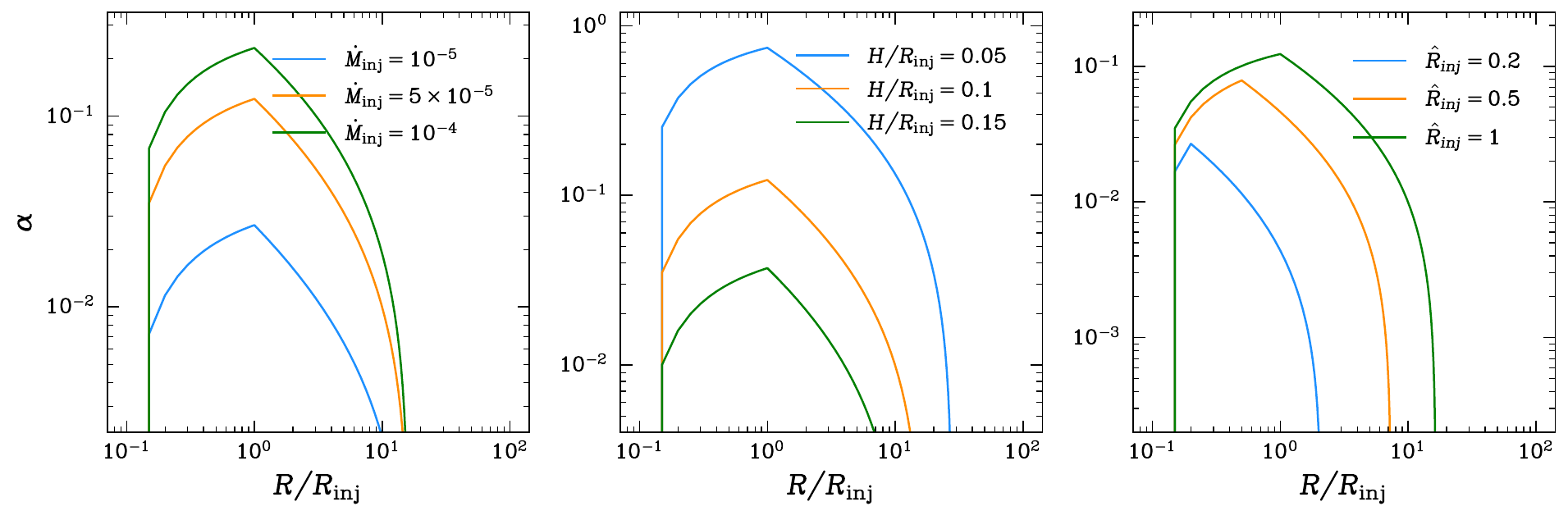}
    \caption{Radial profiles of the $\alpha-$viscosity for different injection rates (left panel), aspect ratios (central panel) and injection radius (right panel) at $t=2t_{\rm inj}$.}\label{fig:alphas}
\end{figure*}

\begin{figure*}
    \includegraphics[width = \textwidth]{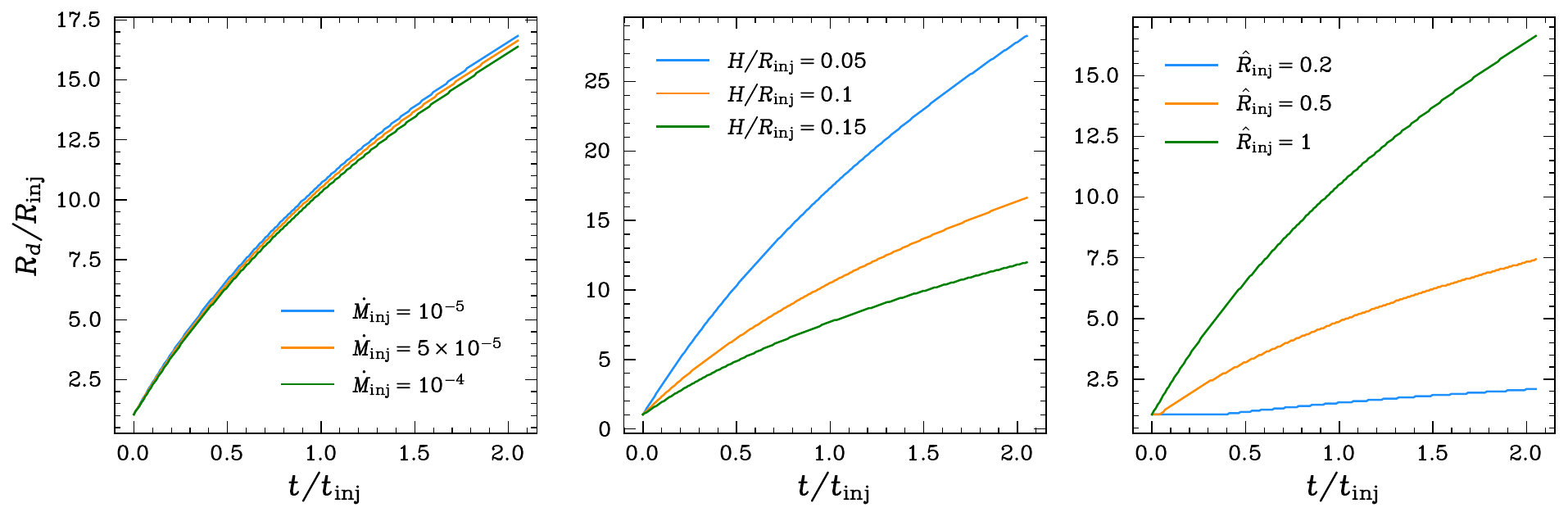}
    \includegraphics[width = \textwidth]{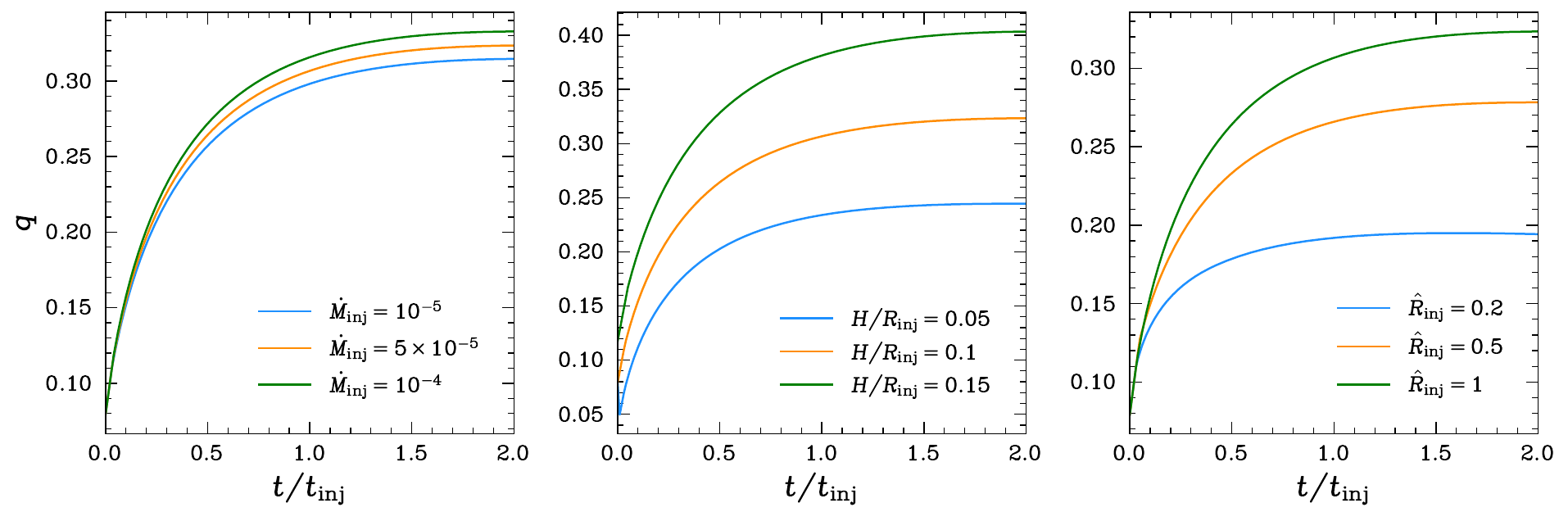}
    \caption{\textbf{Top row:} Time evolution of the outer disc radius for different injection rates (left panel), aspect ratios (central panel) and injection radius (right panel). \textbf{Bottom row:} Time evolution of the disc to star mass ratio for different injection rates (left panel), mass aspect ratios (central panel) and injection radius (right panel). }\label{fig:radius_params}
\end{figure*}

\begin{figure*}
    \centering 
    \includegraphics[width=0.95\linewidth]{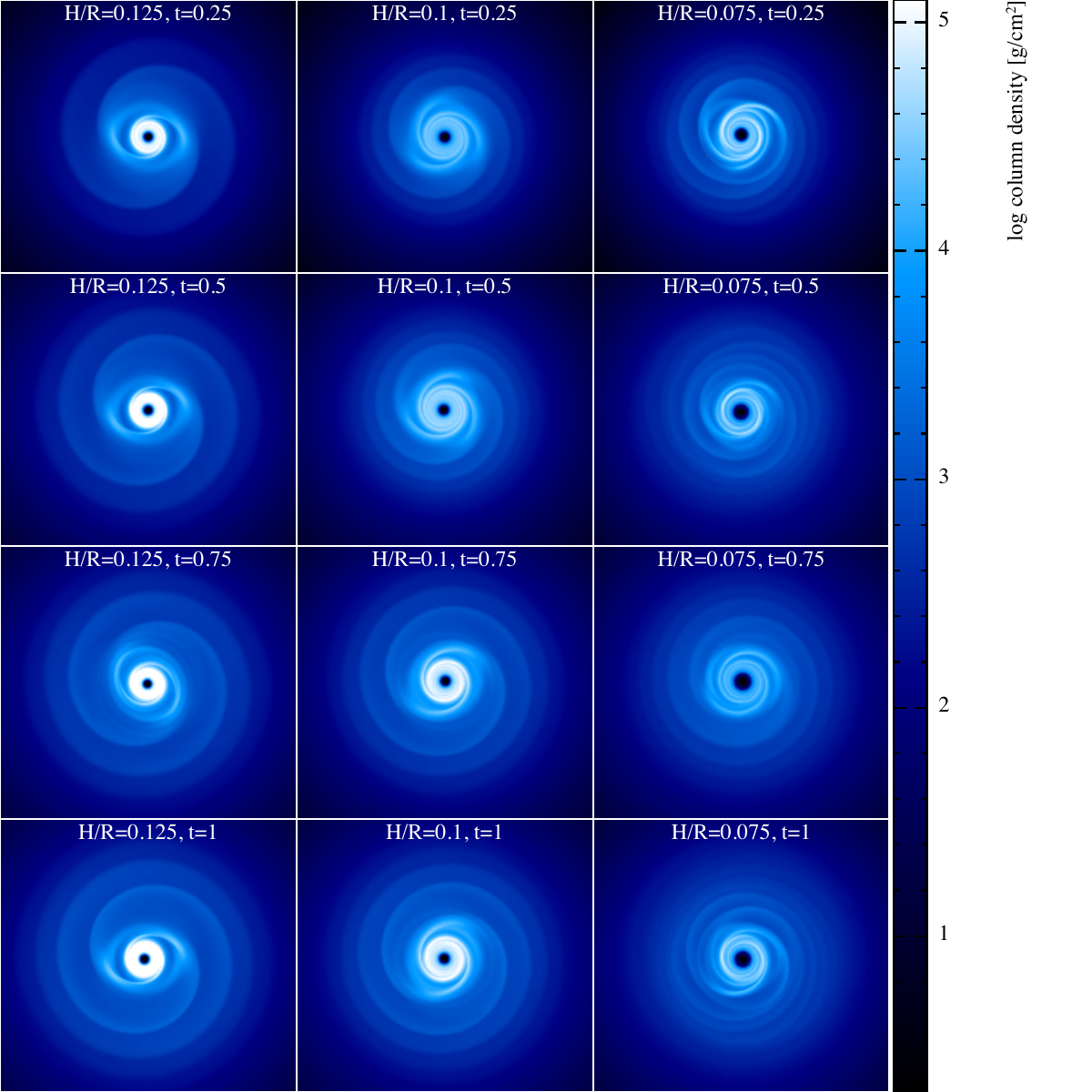}
    \caption{\textit{3D SPH simulations of infall driven gravitational instability.} Snapshots of the logarithmic surface density of SPH simulations with $H/R=0.05$ (left column), $H/R=0.1$ (left column) and $H/R=0.15$ (left column) at $t=0.25t_{\rm inj}$ (first row), $t=0.5t_{\rm inj}$ (second row), $t=0.75t_{\rm inj}$ (third row), $t=t_{\rm inj}$ (fourth row).}
    \label{fig:collage_simulations}
\end{figure*}

\subsection{1D simulations}
Figure~\ref{fig:G1_img1} shows the time evolution of the stellar mass, disc mass and disc to star mass ratio in the 1D evolution code for the reference simulation $\textbf{S1D\_1}$, where time is given in units of injection time. Both the star and disc mass increase with time, eventually reaching a constant disc to star mass ratio. This is particularly evident in the time derivative of these quantities (right panel of Figure~\ref{fig:G1_img1}). The time needed to achieve self-regulation is of the order of the injection time. In general, we observe that this trend is respected in all the simulations. The right panel of Figure~\ref{fig:G1_img1} also shows that $\dot{M}_\star$ and $\dot{M}_d$ are in line with the expected values given by Eqs.~\eqref{mdot_star_SG} and \eqref{mdot_disc_SG}, where $q$ is chosen as the final one, that are the dashed lines. 

Figure~\ref{fig:collection} shows the evolution of the surface density, kinematic viscosity, $\alpha-$viscosity and Toomre $Q$ parameter. As a consequence of the mass injection, the disc mass increases and hence $Q$ decreases. When the $Q=1$ threshold is hit, the viscosity switches on and the angular momentum is transported throughout the disc. Eq.~\eqref{alpha_TH} describes how the $\alpha$-viscosity depends on the injection rate and aspect ratio: $\alpha$ is proportional to the injection rate, and a thinner disc results in more efficient angular momentum transport. Fig.~\ref{fig:alphas} shows radial profiles of the $\alpha-$viscosity for different injection rates (left panel), aspect ratios (central panel) and injection radius (right panel) at $t=2t_{\rm inj}$. The profiles peak at the injection radius, where the mass is injected and hence the Toomre parameter reaches its minimum. The trends with $\dot{M}_{\rm inj}$ and $H/R$ are recovered as expected from 
Eq.~\eqref{alpha_TH}. As for the relationship between $\alpha$ and $\hat{R}_{\rm inj}$, the viscosity is ultimately determined by the value of $\Omega$ at the injection radius. Since the Keplerian frequency decreases with radius, we expect a larger  $\alpha$ for a greater $\hat{R}_{\rm inj}$, as observed in Figure~\ref{fig:alphas}.

As a result of angular momentum transport, the disc undergoes viscous expansion, as illustrated in the top-left panel of Figure~\ref{fig:collection}. The extent of this spreading depends on the viscosity, which is influenced by the aspect ratio, mass injection rate, and injection radius. The top row of Figure~\ref{fig:radius_params} shows the time evolution of the disc's outer radius as a function of the injection rate (left panel), aspect ratio (central panel), and injection radius (right panel). These trends reflect the dependence of viscosity on these parameters.

As described in the previous section, the self-regulated state is characterised by having a constant disc to star mass ratio $q$, that ensures that the Toomre parameter $Q\approx 1$ throughout the disc. The bottom row of Figure~\ref{fig:radius_params} shows how the self-regulated $q$ depends on the simulation parameters. Although the injection rate does not have much impact on the equilibrium $q$, the aspect ratio does, as it determines the amount of mass needed to reach the $Q=1$ state. Thicker, and hotter, discs show a higher disc to star mass ratio, as expected. In addition, the location of the injection plays a role in determining the equilibrium disc to star mass ratio, because it determines the region of the disc that must be in a $Q=1$ state, to transport angular momentum. 

The 1D simulations are useful to understand the main trends, since they are computationally cheap and easy to interpret, however they do not capture the complexity of the full three-dimensional simulations. Indeed, with the 1D evolution code it is not possible to simulate disc fragmentation due to GI, that is known to happen in a region of the parameter space \citep{kratter10a,kratter16}.

\subsection{SPH simulations}
Figure~\ref{fig:collage_simulations} shows shapshots of column density evolution in the 3D \textsc{Phantom} simulations, for three different disc aspect ratios (left to right) at four different times (top to bottom), where time is given in units of the injection timescale (Eq.~\ref{eq:tinj}). Thinner discs (right column) show more spiral arms because the saturation value of the disc-to-star mass ratio is lower, and according to the cubic dispersion relation of \citet{lau78}, the most unstable mode moves to higher $m$ in discs with lower $q$. This also explains the trend with time, as $q$ increases with time towards its saturation value (top to bottom), also producing more spiral arms.

\begin{figure}
    \centering
    \includegraphics[width=\linewidth]{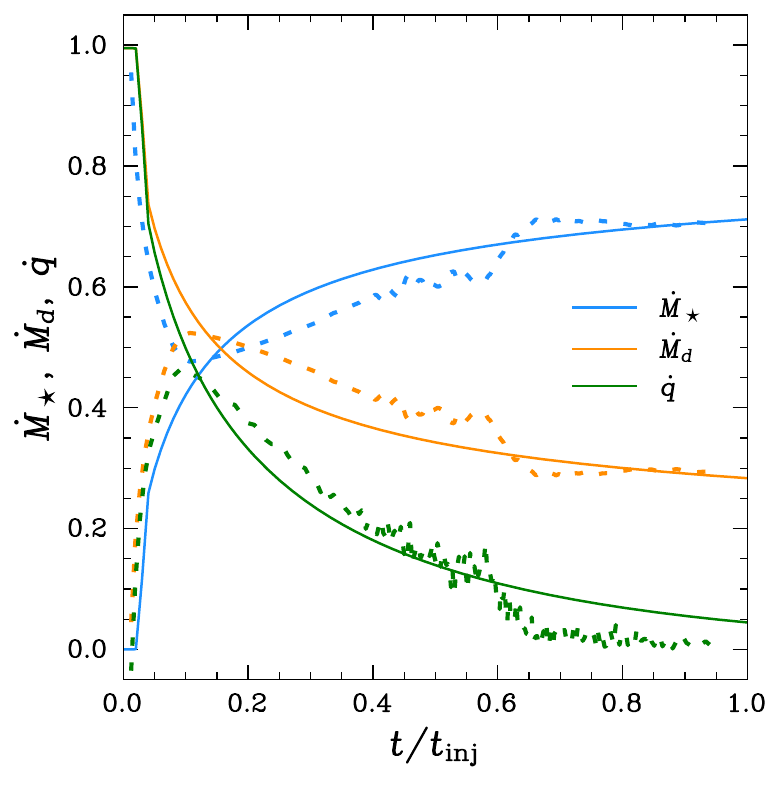}
    \caption{Comparison between the 1D evolution code and the 3D SPH simulation. The solid lines correspond to the evolution of  $\dot{M}_\star$ (blue), $\dot{M}_d$ (orange) and $\dot{q}$ (green) according to the 1D evolution code, for the reference simulation. The dotted lines show the behaviour of the reference 3D SPH simulation, with the same initial parameters.}
    \label{fig:comparison}
\end{figure}

\subsubsection{Comparison with 1D code}

Figure~\ref{fig:comparison} compares the evolution of disc mass, star mass, and disc-to-star mass ratio between the 1D and 3D reference simulations ($\textbf{S1D\_1}$ and $\textbf{S3D\_1}$). Apart from the dimensionality, the key physical difference between the 1D and 3D simulations is the viscosity. Indeed in the 1D case the value of $\alpha$ is a prescribed function of $Q$ which self-regulates so as to achieve the required steady state $\alpha$; in the 3D case conversely the development of the instability is modelled hydrodynamically and the value of $\alpha$ quantifies  angular momentum transport associated with spiral structures.

The results of the \textsc{Phantom} simulation appear to be well described by the simple 1D code. In 3D the system also relaxes into a self-regulated state with a constant disc to star mass ratio. In addition, the final values of the accretion rates are in line with the analytical expectations. Table~\ref{table_3D_global} compares the results of the 3D simulations to the theoretical expectations, showing agreement to $\lesssim 10$\% between the measured and expected mass growth rates.

\begin{table}\centering\caption{Global quantities of the 3D simulations at self-regulation compared with the analytical expectations. The star and disc accretion rates are in units of the injection rate.}
{\begin{tabular}{lllllll}
 $\dot{M}_{\rm inj}$ & $H/R$ & $q$ & $\dot{M}_\star$  & $\dot{M}_d$ & Exp. $\dot{M}_\star$  & Exp. $\dot{M}_d$ \\ \hline
$2.5\times10^{-5}$  &  $0.075$& $0.25$  & $0.75$  & $0.25$ & $0.8$  & $0.2$\\
$5\times10^{-5}$  &  $0.1$& $0.32$  & $0.7$  & $0.3$ & $0.75$  & $0.25$\\
$1\times10^{-4}$  &  $0.125$& $0.39$  & $0.7$  & $0.3$ & $0.71$  & $0.29$\\

\end{tabular}}\label{table_3D_global}
\end{table}
 
\subsubsection{Azimuthal modes analysis}

\begin{figure}
    \centering
    \includegraphics[width=0.95\linewidth]{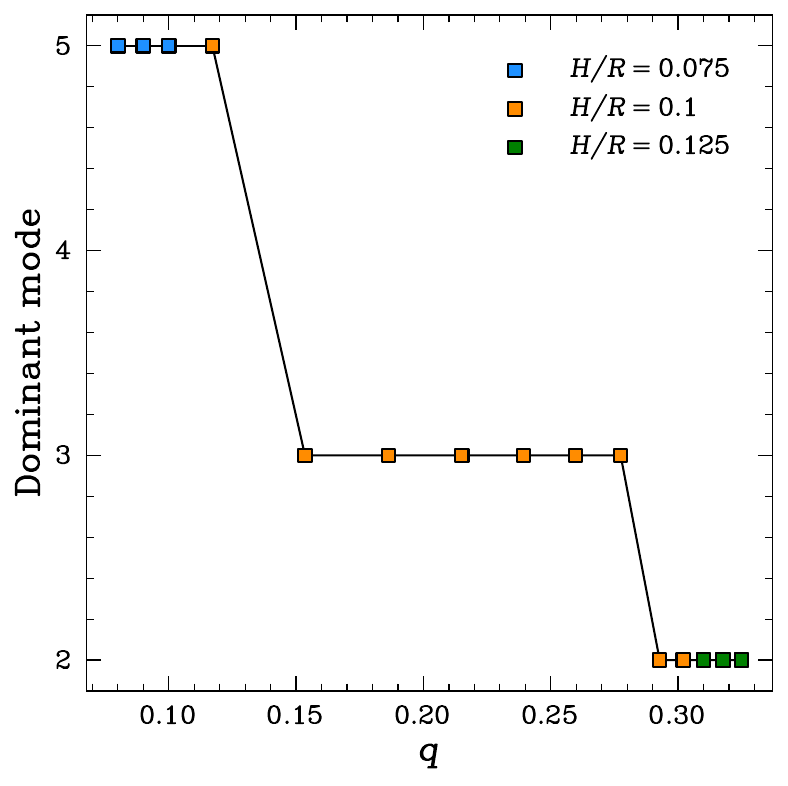}
    \caption{Dominant azimuthal mode as a function of the disc to star mass ratio, combining the three \textsc{Phantom} simulations.}
    \label{fig:dominant_m}
\end{figure}

\begin{figure*}
    \centering
    \hspace{0.7cm}\includegraphics[width=0.625\linewidth]{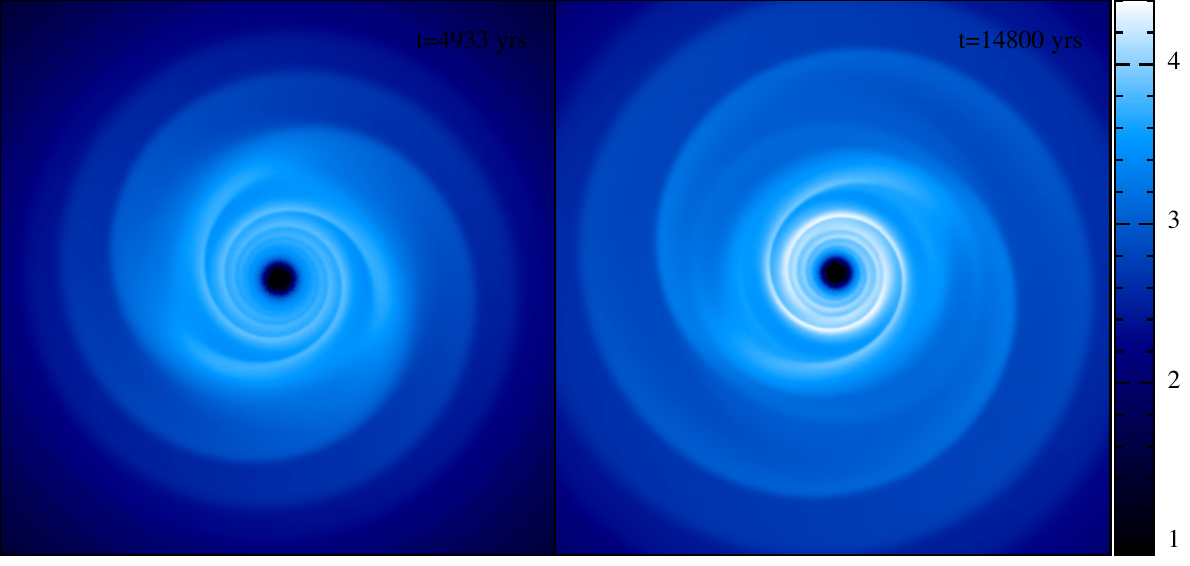}
    
    \hspace{-0.8cm}\includegraphics[width=0.645\linewidth]{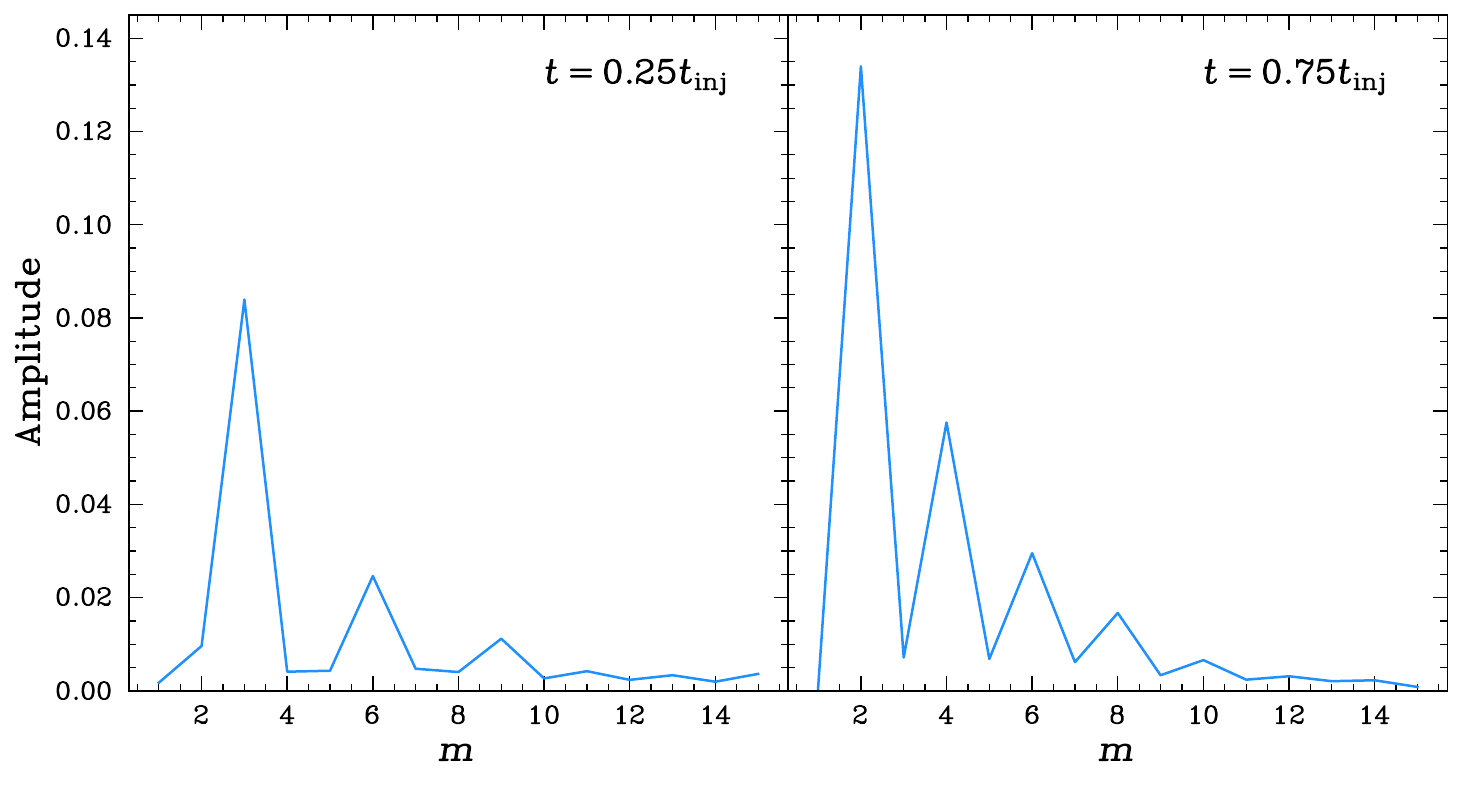}
    \caption{Density snapshots and the corresponding power spectrum for the reference simulation at $t=0.25$ and $t=0.75$ at $R=7.5$.}
    \label{fig:pspec_time}
\end{figure*}

\begin{figure}
    \centering
    \includegraphics[width=0.95\linewidth]{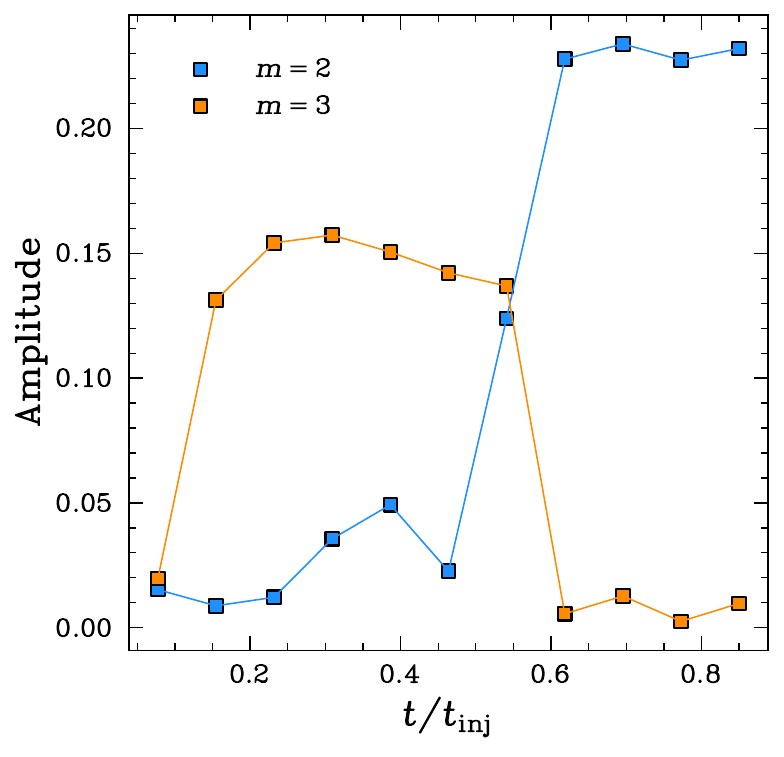}
    \caption{Evolution of the $m=2$ and $m=3$ modes at $R=7.5$ in the reference simulation as function of time.}
    \label{fig:m2m3}
\end{figure}

We start the morphological characterisation by studying the azimuthal modes of the perturbation. We extract the azimuthal wavenumber $m$, i.e. the number of spiral arms, using the procedure described in \cite{cossins09}.  We first divide the disc into radial annuli with fixed width $\Delta R = 0.2$. Then, within each annulus, the azimuthal wavenumber amplitude $\mathcal{A}_m$ is computed directly from the SPH particles' position according to 
\begin{equation}
    \mathcal{A}_m = \frac{1}{N_{\rm ann}}\left|\sum_{i=1}^{N_{\rm ann}}\exp[-im\phi_i]\right|,
\end{equation}
where $\phi_i$ are the azimuthal angles of the individual particles and $N_{\rm ann}$ is the number of particles within an annulus.

The number of spiral arms is determined by the disc to star mass ratio: in particular, high disc mass discs show fewer spiral arms \citep{lau78, laughlin94}. 
Figure~\ref{fig:dominant_m} shows the dominant azimuthal wavenumber as a function of the disc to star mass ratio. To produce the plot, we combined the results of the three \textsc{Phantom} simulations, using the fact that $q$ increases with time. As expected, a thicker disc (higher $q$) has a lower dominant $m$ mode.

This behaviour is clearly seen in Figure~\ref{fig:pspec_time}, which shows the morphology (top row) and power spectrum (bottom row) of the reference simulation at two distinct time steps, $t=0.25$ (left) and $t=0.75$ (right) at a radius $R=7.5$. At $t=0.25$, the disc has not yet reached self-regulation, and its disc-to-star mass ratio is still increasing. Consequently, \( q(t=0.25) < q(t=0.75) \), leading to a different spiral morphology. At this stage, the dominant mode is $m = 3$, as evidenced by the power spectrum peaks occurring at multiples of three. In contrast, by \( t = 0.75 \), the disc has achieved self-regulation and the mass ratio has stabilized to a level where the dominant mode shifts to \( m = 2 \), with peaks of the power spectrum appearing in multiples of two. As  mentioned above, the disc to star mass ratio controls the azimuthal wavenumber. 

Figure~\ref{fig:m2m3} illustrates the transition between $m=2$ and $m=3$ modes more clearly, by plotting the amplitude of these modes at $R=7.5$ in the disc as a function of time. A sharp transition from $m=3$ to $m=2$ modes can be seen to occur, controlled by the disc to star mass ratio. Comparison with Fig.~\ref{fig:comparison} shows that this transition happens when self-regulation kicks in, at approximately $t=0.5t_{\rm inj}$.

\subsubsection{Morphological differences with cooling driven GI}
\label{sec:morphology}
\begin{figure*}
    \centering
    \hspace{1.55cm}\includegraphics[width=0.6275\linewidth]{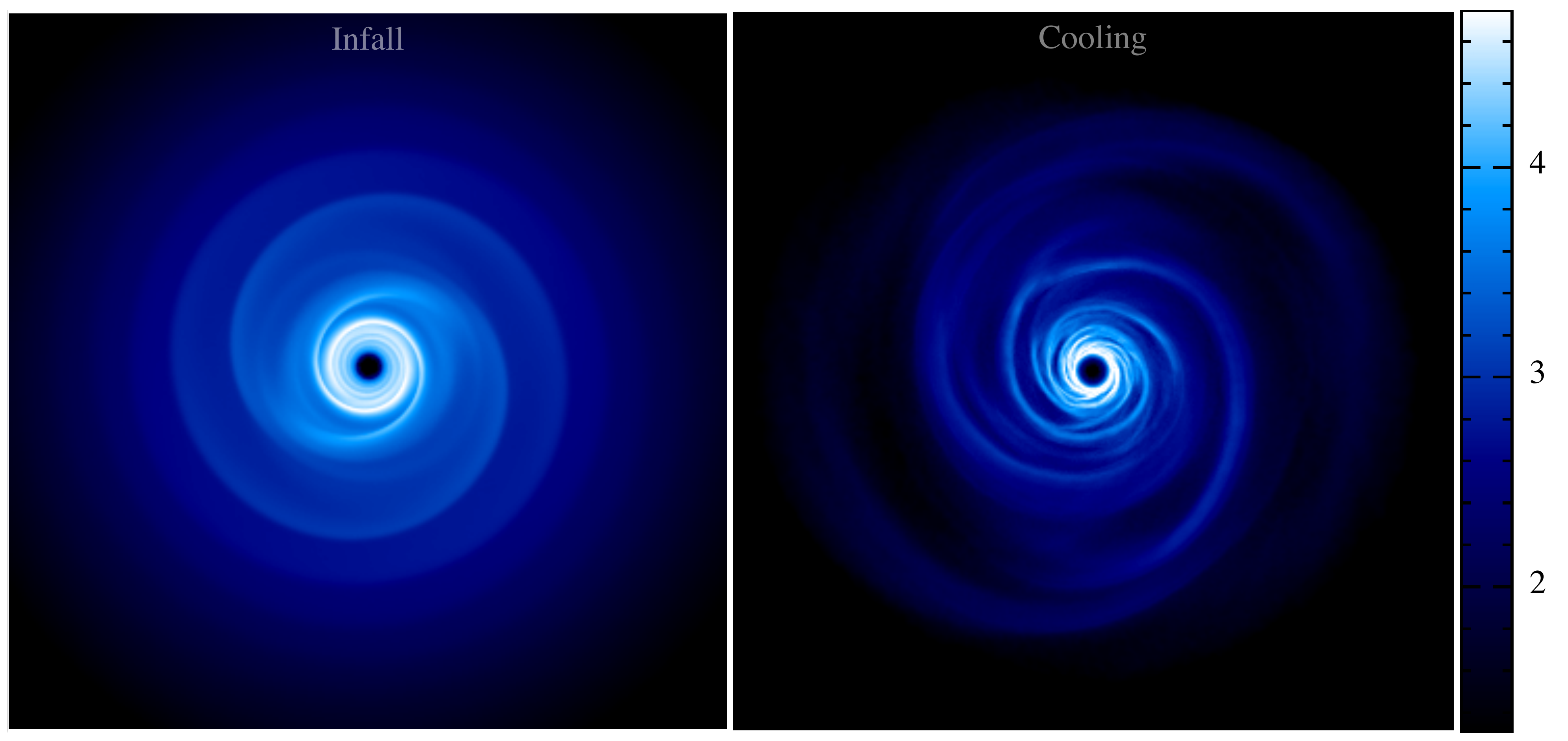}
      \hspace{-2cm}\includegraphics[width=0.6457\linewidth]{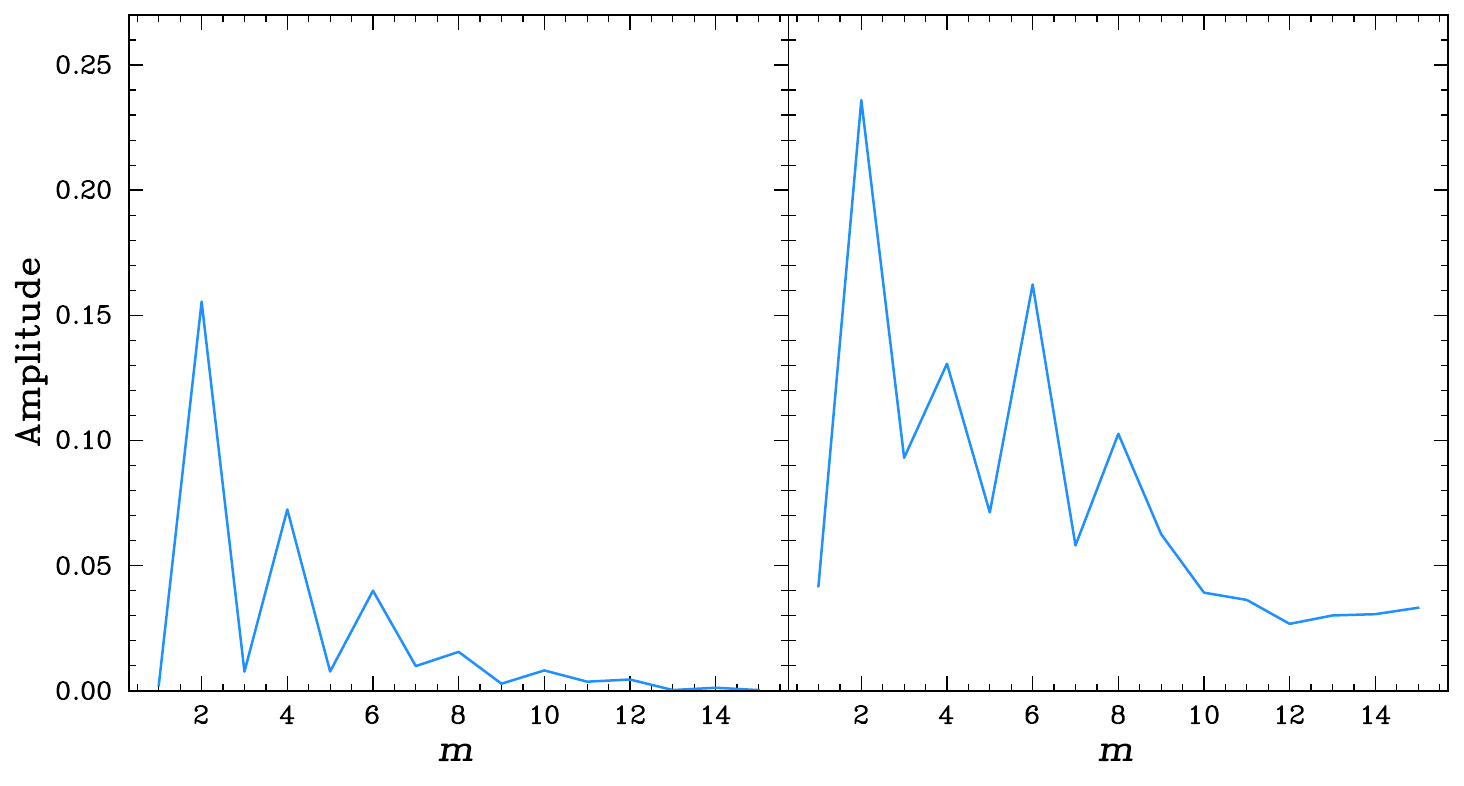}
    \caption{Surface density snapshots (top row) for the reference simulation (left) and the cooling driven (right), and the corresponding azimuthal power spectrum at $R=5$ (lower panels).}
    \label{fig:cool_inf}
\end{figure*}

Figure~\ref{fig:cool_inf} compares the reference simulation with its counterpart where the GI is triggered by cooling, both having the same final disc-to-star mass ratio. From the density snapshots, we observe that the dominant mode remains the same, \( m = 2 \), since it is just determined by the disc to star mass ratio. Yet important morphological differences emerge. In the cooling simulation, the pitch angle remains nearly constant with radius, whereas in the infall case, as we will show below, this is not the case. More importantly, the cooling-driven spirals are more flocculent, exhibiting transient secondary spiral arms that continuously form and dissipate. This is particularly evident in the azimuthal power spectra shown in the bottom panel of Figure~\ref{fig:cool_inf}. While both simulations display a peak at \( m = 2 \), indicating the dominant mode, the cooling simulation exhibits power across higher \( m \) modes, emphasising its flocculent nature.

%The main driver of this change in morphology is the differing temperature profile of the disc in each case. In the infall-driven simulation, the temperature remains fixed due to the locally isothermal assumption, meaning that spiral structures do not heat the disc, and surface density is the only parameter that can vary. In contrast, in the cooling-driven simulation, both surface density and temperature evolve, leading to the formation of secondary spiral arms and contributing to the flocculent appearance.

{We believe that there are two main drivers of the differences between the cooling and the infall-driven cases. The first is the difference in thermodynamic treatment: the cooling simulations solve the full energy equation and allow the temperature to evolve, while the infall-driven simulations assume a locally isothermal equation of state. This certainly affects the dynamics and the development of instabilities. However, it is worth noting that previous studies using a variety of thermodynamic and infall prescriptions — including \cite{kratter08}, \cite{zhu12}, and \cite{harsono11} — consistently report the emergence of coherent spiral structures, often resembling global modes excited near the infall radius. For instance, \cite{kratter10a} and \cite{zhu12} adopt different treatments of the thermal physics (with the former using a locally isothermal approximation and the latter implementing radiative cooling), yet both identify similar morphological features to those found in our simulations. \cite{harsono11}, on the other hand, include infall onto a disk governed by beta cooling. The resulting morphology exhibits a more coherent global mode structure, albeit with residual flocculent spirals characteristic of beta-cooled discs. These comparisons suggest that the presence of coherent spiral patterns may be a robust outcome of infall, somewhat independent of the exact thermodynamic treatment.
Secondly, in the infall-driven case, the angular momentum of the infalling material plays a crucial role in determining the modes that are excited and the onset of the non-linear evolution. In this paper, we have considered an idealised case where the infalling material has a fixed angular momentum, exactly the Keplerian angular momentum at the injection radius. This excites a coherent global mode with a pattern speed matching the Keplerian value at that radius. In more realistic scenarios, the angular momentum of the infalling material would span a broader range, potentially exciting a richer spectrum of modes.}

\subsubsection{Spiral tracking and pattern speed}

\begin{figure*}
    \centering
    \includegraphics[width=0.495\linewidth]{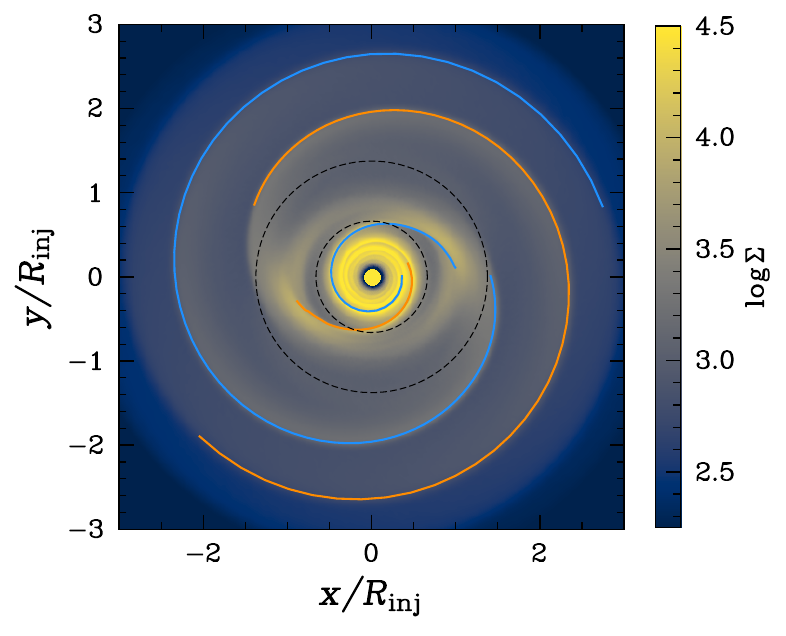}
    \includegraphics[width=0.4\linewidth]{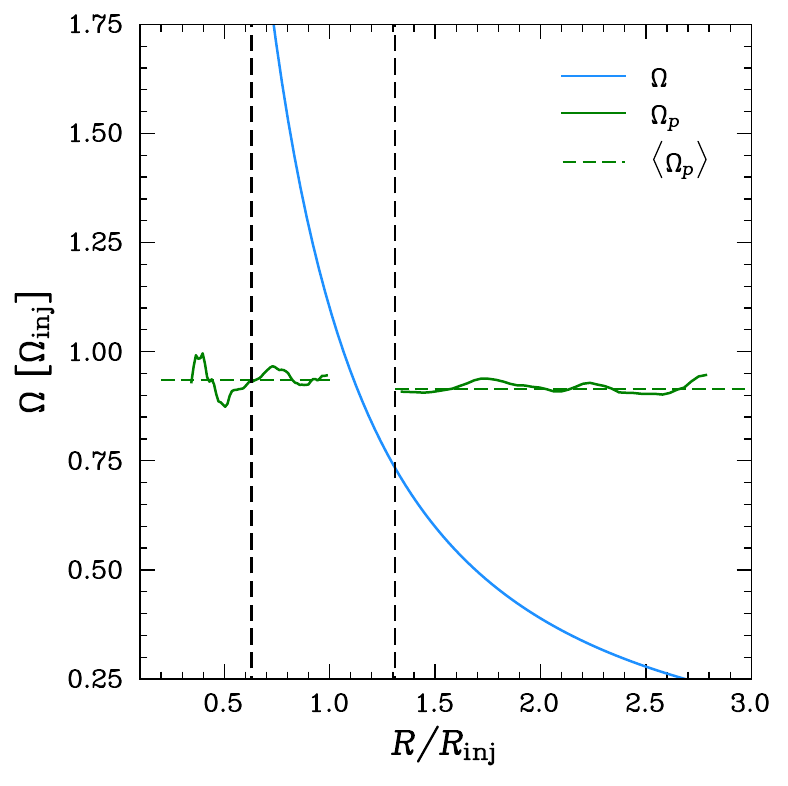}
    \caption{Left panel: spiral arms identification procedure for the \textbf{S3D\_3} simulation after 1 injection time. The dominant mode is $m=2$, and the two spiral arms are underlined with blue and orange lines. Right panel: pattern speed of the identified spirals (green solid line), radially averaged value (green dashed line) and $\Omega$ extracted from the simulation (blue line). {The black dashed lines correspond to the inner and outer Lindblad resonances.}}
    \label{fig:spiral_pitch_k}
\end{figure*}

\begin{figure*}
    \centering
    \includegraphics[width=0.925\linewidth]{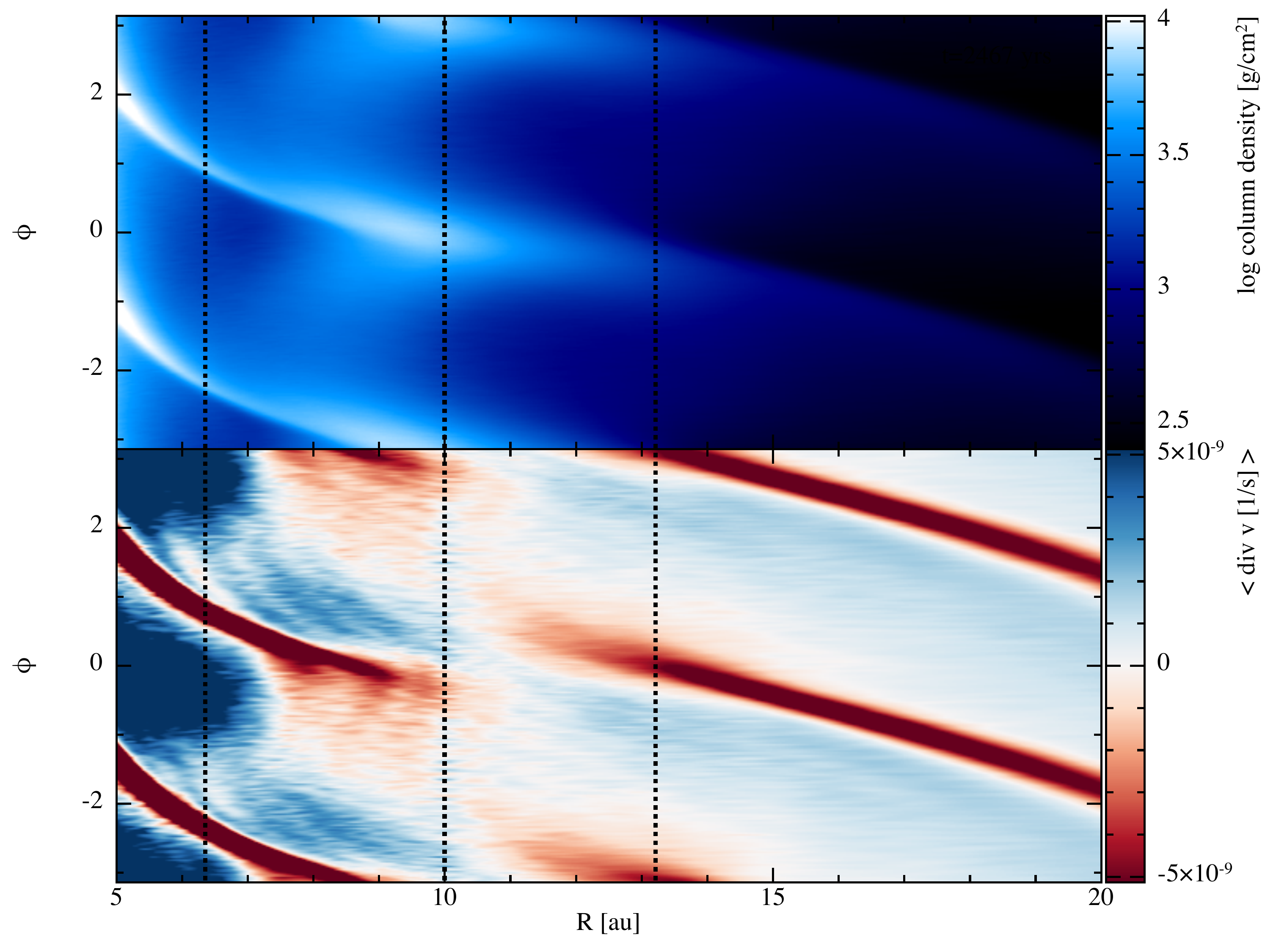}
    \caption{Polar plot of the surface density $\Sigma$ (top panel) and of the divergence of the velocity field $\nabla\mathbf\cdot{v}$ (bottom panel) of the \textbf{S3D\_3} simulation after 1 injection time.}
    \label{fig:spiral_breaking}
\end{figure*}

\begin{figure*}
    \centering
    \includegraphics[width=0.475\linewidth]{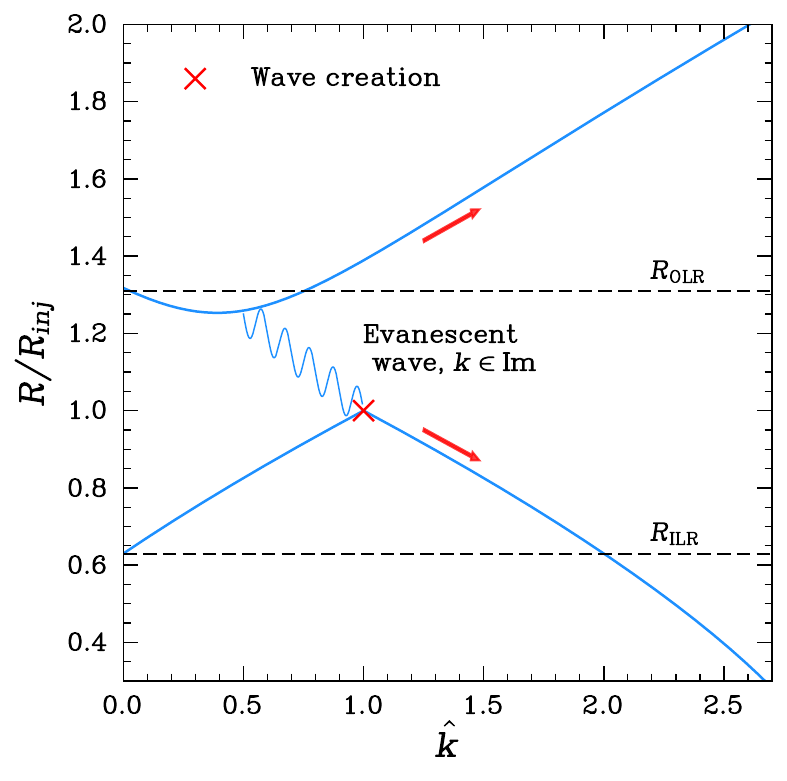}  
    \includegraphics[width=0.475\linewidth]{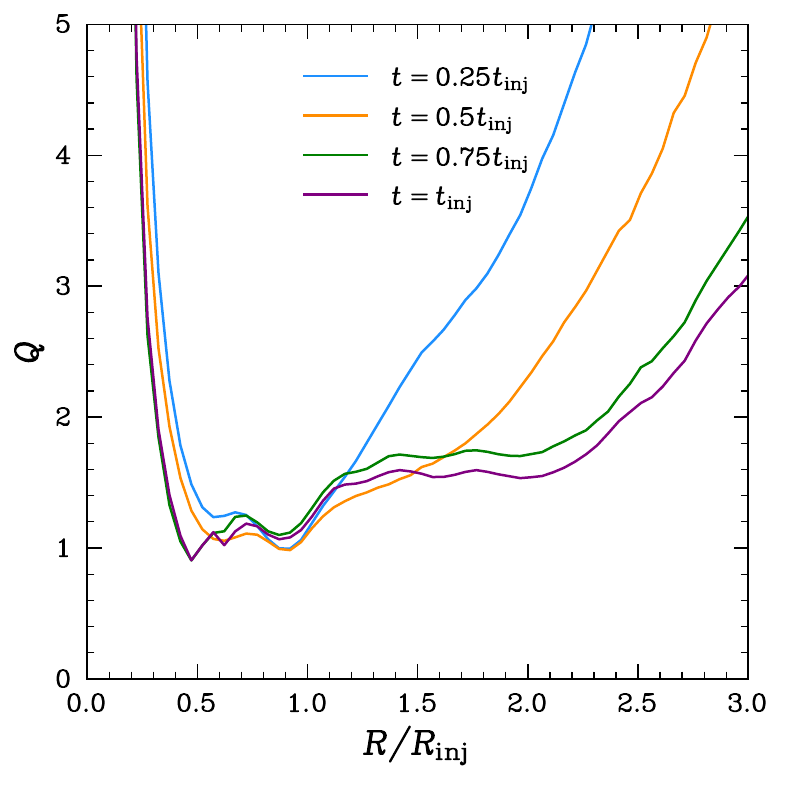}
    \caption{Left panel: Propagation diagram of trailing waves in a Keplerian disc, according to Eq. \eqref{DR_kHR}. The waves are generated at the corotation-injection radius (red cross) and they propagate inwards-outwards following the short trailing branches (red arrows). The outward wave ''tunnels'' through the evanescent zone, where $k\in$ Im. Right panel: Toomre parameter of the \textbf{S3D\_3} simulation as a function of radius for different times.}
    \label{fig:wave_propagation}
\end{figure*}

In order to perform the radial mode analysis and to calculate the pattern speed of the perturbation in the infall-driven scenario, we track the spiral arms. To identify the spiral arms in the simulations, we find the peaks in the radial derivative of the surface density\footnote{We tried to track the spiral arms both from the maxima in the surface density and in the radial derivative of $\Sigma$, and we found that the latter quantity better trace the perturbations.} using \verb+find_peaks+ implemented in the Python library \textsc{scipy} \citep{2020SciPy-NMeth}. The left panel of Fig. \ref{fig:spiral_pitch_k} shows the tracking procedure for the \textbf{S3D\_3} simulation after 1 injection time. The tracking procedure allows us to compute the pattern speed of the spiral, comparing the position of the maxima at different times. The right panel of Fig. \ref{fig:spiral_pitch_k} shows the pattern speed of the spirals as a function of radius, compared with $\Omega(R)$ extracted from the simulation. The pattern speed is constant throughout the radial extent of the disc, marking a fundamental difference from the cooling-driven case. {Indeed, in the standard GI scenario \citep{cossins09}, the instantaneous pattern speed of the spiral is close to the local Keplerian frequency. Spiral density waves are typically generated near corotation, and shock quickly as they propagate away from it, dissipating.} 

Interestingly, in the infall scenario $\Omega_p$ matches the angular frequency at the injection radius $\Omega_{\rm inj}$, and this recovered in every simulation, as shown in Appendix \ref{app:tracking}. This behaviour may be interpreted as the infall triggering a spiral mode with a pattern speed that matches the velocity of the injected material. 

Since the pattern speed of the perturbation is constant with radius, it is possible to compute the Lindblad resonances according to eq. \eqref{lindblad_resonances_keplerian}. In this case $m=2$, and the resonances are at $R_{\rm ILR} = 0.63R_{\rm inj}$ and $R_{\rm OLR} = 1.31R_{\rm inj}$. Clearly, the corotation resonance occurs at the injection radius. In the following, the Lindblad resonances are marked on the figures with black dashed lines. 

By visually inspecting the spiral morphology, as shown in Fig. \ref{fig:spiral_breaking}, something interesting happens between the injection radius and the outer Lindblad resonance. The bottom panel of Fig. \ref{fig:spiral_breaking} shows the divergence of the velocity field $\nabla v$, a quantity that tracks the spirals. Indeed, when $\nabla v<0$ the flow is converging, and that happens at the wave front. The spiral structure clearly breaks down between the injection radius and the outer Lindblad resonance. To understand the cause of this behaviour, we need to analyse wave propagation in that region using the dispersion relation.

The dispersion relation can be written in dimensionless units according to
\begin{equation}\label{dless_disprel}
    s^2 = {Q^2}\hat{k}^2 - 2|\hat{k}| +1,
\end{equation}
where $s=m (\Omega_p-\Omega)/\kappa$ and $\hat{k} = k / (\kappa^2 / \pi G \Sigma)$. This variable is particularly useful because $s=0$ corresponds to corotation, and $s=\pm1$ to the Lindblad resonances. Under the assumption of Keplerian potential, which is justified in our simulations, the variable $s$ can be re-written as
\begin{equation}
    s = m\left(\hat{R}^{3/2}-1\right),
\end{equation}
where $\hat{R}$ is the radius in units of the corotation one, that is the injection radius in this case. Combining this with the dispersion relation, we can obtain how kH changes with radius, depending on the Toomre parameter $Q$
\begin{equation}\label{DR_kHR}
    \hat{R}=\left[\pm\frac{1}{m}\sqrt{{Q^2}\hat{k}^2 -2|\hat{k}|+1} +1    \right]^{2/3}.
\end{equation}
As shown in \citet{binneytremaine}, Chapter 6, there are four families of possible solutions:
\begin{itemize}
    \item Short trailing waves ($k>0$ and $kH$ increasing with $|s|$),
    \item Long trailing waves ($k>0$ and $kH$ decreasing with $|s|$),
    \item Short leading waves ($k<0$ and $kH$ increasing with $|s|$),
    \item Long leading waves ($k<0$ and $kH$ decreasing with $|s|$).
\end{itemize}
In this analysis we are interested in trailing waves ($k>0$), so we can discard the leading solutions. The left panel of Fig. \ref{fig:wave_propagation} shows the dispersion relation in the $(\hat{R},kH)$ plane, {where we fixed $m=2$}, using the Toomre parameter extracted from the simulation, shown in the right panel of Fig. \ref{fig:wave_propagation}, i.e. $Q\sim1$ for $\hat{R}<1$ and $Q\sim1.5$ for $\hat{R}>1$. Since the Toomre parameter is equal to 1 inside the corotation, a wave that is created at the injection radius can propagate inward following the short trailing branch (decreasing the pitch angle far from the injection), $k$ being a real number. Conversely, since $Q>1$ outside the corotation, the radial wavenumber $k$ is imaginary, hence the wave is evanescent and its amplitude decreases. At some point, far out enough from the injection radius, the radial wavenumber becomes real again, and the wave can propagate, following again the short trailing branch. It happens that the location where $k$ becomes real is close to the outer Lindblad resonance for $Q\sim1.5$, which matches what we see in the simulation (see Fig. \ref{fig:spiral_breaking}).

This argument explains why the spirals break between the injection radius and the outer Lindblad resonance\footnote{Formally, the propagation of the wave across the evanescent region is analogous to the quantum mechanical tunneling effect. In both cases, the wavenumber becomes imaginary, leading to an exponential decay of the wave amplitude. However, it is important to stress that this is only a mathematical analogy: no actual quantum effects are involved.}. Hence, in that region we are not able to track them, and in the following plot that region will be excluded from our analysis. 

We stress that the analysis presented here on wave propagation is necessarily simplified. A more rigorous treatment would require the use of higher-order dispersion relations \citep{lau78}, or ideally a global stability analysis that properly accounts for boundary conditions and normal modes \citep{shu70a,shu70b} and allows one to compute growth rates of the perturbations. Such an approach would provide a more complete understanding of the nature and evolution of the modes triggered by the infall. We leave this detailed investigation to a forthcoming paper.

\begin{figure*}
    \includegraphics[width=0.475\linewidth]{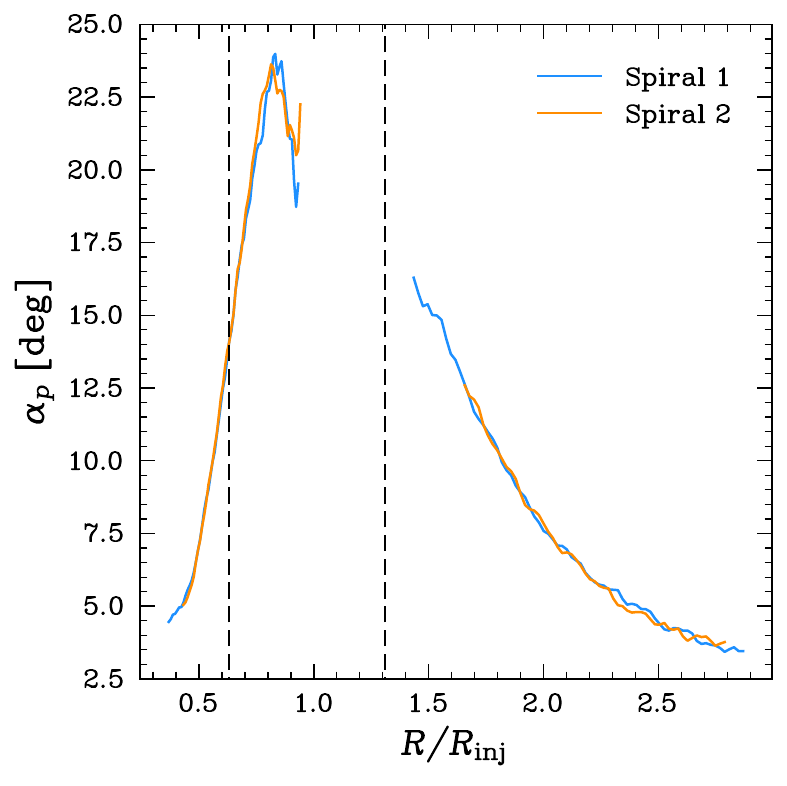}
    \includegraphics[width=0.475\linewidth]{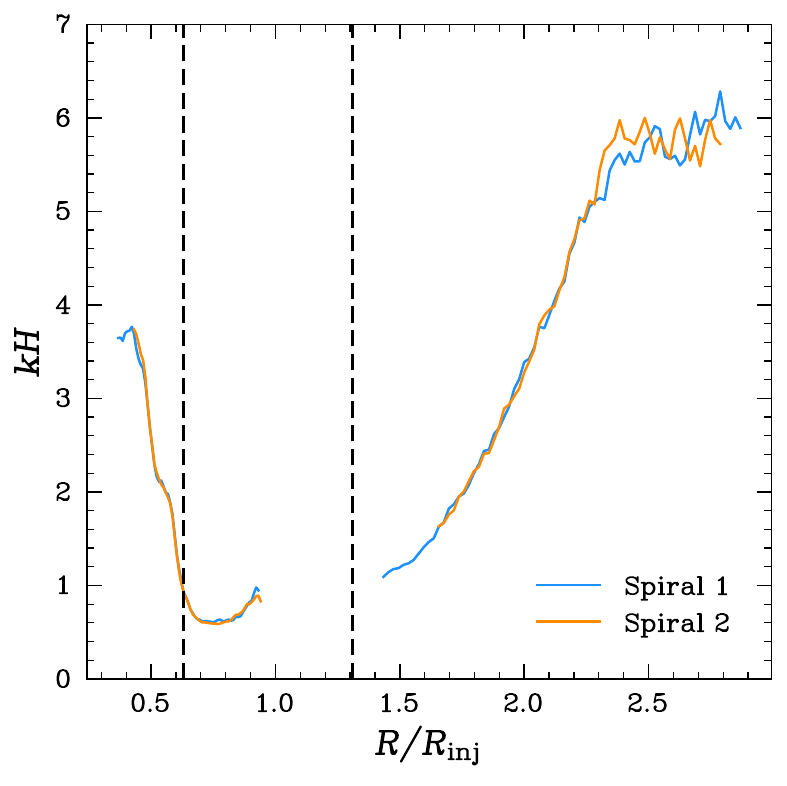}
    \caption{Left panel: Pitch angle of the \textbf{S3D\_3} simulation after 1 injection time computed using eq. \eqref{pitch_eq}. Right panel: $kH$ profile  of the \textbf{S3D\_3} simulation after 1 injection time. This quantity increases with distance from corotation, pointing to the short-trailing nature of the waves.}
    \label{fig:pitch_KH}
\end{figure*}

\subsubsection{Radial mode analysis}
Performing the radial mode analysis is more challenging compared to the azimuthal modes. As pointed out by \cite{cossins09}, radial binning of the data is crucial in order to obtain the correct result. We tried to use the same procedure of \cite{cossins09}, but we found that our results strongly depended on the radial binning, and did {not} converge. For this reason, we instead derive the radial wavenumber $k$ by measuring the pitch angle of the spiral, since they can be related according to 
\begin{equation}\label{k_wvnum}
    k = \frac{m}{R\tan \alpha_p}.
\end{equation}
To compute the opening angle $\alpha_p(R)$ we leverage on the tracking procedure outlined before. We compute the opening angle of the identified spirals through geometrical arguments, where $\alpha_p$ represents the angle between the tangent to the spiral arm and the tangent to a circular arc at a given radius, according to
\begin{equation}\label{pitch_eq}
    \tan\alpha_p = \left|\frac{1}{R}\frac{\text{d}R}{\text{d}\phi} \right|.
\end{equation}
In principle, each identified spiral arm has a distinct $k$; however, we verify that the individual radial wavenumbers are consistent and take the average of $k$ across all identified spirals for the purposes of our analysis.

The right panel of Fig.~\ref{fig:pitch_KH} shows the pitch angle of the \textbf{S3D\_3} simulation after 1 injection time. Again, the Lindblad resonances are shown as black dashed lines. The pitch angle profile shows a maximum at $R\sim R_{\rm inj}$ with smaller opening angles both inside and outside this radius. This behaviour resembles the planetary case, where the wave propagation is almost radial at the planetary location, and then the perturbations become tightly wound \citep[e.g.][]{ogilvie02,rafikov02}. From the pitch angle we then compute the dimensionless radial wavenumber $kH$ using Eq.~\eqref{k_wvnum}, where we define $H$ as\footnote{{We underline that for a gravitationally unstable disc, this definition of $H$ is equivalent to $\hat{k}/k$, as used in eq.\eqref{dless_disprel}.}}
\begin{equation}
    H=\frac{c_s^2}{\pi G \Sigma}.
\end{equation}
The right panel of Fig.~\ref{fig:pitch_KH} shows $kH$ as a function of radius. This quantity increases with distance from the injection, confirming the short-trailing nature of these spirals.

\section{Discussion}\label{section5}

\subsection{Cooling or infall?}
\label{sec:thermostat}

\citet{gammie01} showed that the non-linear outcome of gravitational instability in cooling accretion discs is either fragmentation or a state of self-regulated turbulence, depending on the cooling rate. This picture assumes that the only source of heat in the disc is due to spiral shocks caused by the gravitational instability itself. 
%If shut off, the disc cools back to the $Q=1$ state. 
Hence, if the cooling rate is long enough, the disc self regulates  as a temperature-controlled thermostat around $Q=1$. %But is this really the case?

However, in a realistic scenario, heating from the central object plays a crucial role in determining the disc’s thermal equilibrium, potentially influencing its ability to reach a self-regulated state. Gravitational instability in irradiated discs is a delicate topic. Simulations of protostellar discs incorporating realistic heating and cooling mechanisms suggest that stellar irradiation primarily governs the disc temperature \citep{rice11,haworth20,young24,rowther24b}, with internal heating from shocks playing only a minor role on a global scale.

Recent results by \citet{rowther24b}, using on-the-fly Monte Carlo radiative transfer, reinforce the idea that protostellar discs are \emph{passive}, their thermal structure dictated by the central star. Similar conclusions were drawn by previous studies \citep{matzner05,krumholz07,meru10,kratter10a}. These are all global simulations, where the large-scale temperature profile is set by stellar irradiation. However, local deviations can occur. For instance, \citet{leedham25}, using 2D shearing {sheet} simulations with external irradiation, found that when a spiral arm crosses, the temperature varies due to local shock heating. This suggests that while the disc’s overall thermal structure is dictated by the star, local heating from spiral shocks may still play a significant role.

Our work adopts the extreme assumption, treating the disc as locally isothermal. The reality likely lies between these two extremes, necessitating live radiative transfer simulations with infall to assess the impact of material injection.

%How then should we interpret gravitational instability in real discs? 
Interestingly, all of the discs we believe are gravitationally unstable, Elias 2-27, \citep{paneque21} AB Aurigae, \citep{tang12} and GM Aur, \citep{schwarz21}, show evidence of infall from the environment. Our numerical experiments in this paper suggest a picture closer to a bathtub than a thermostat: the disc fills by adding mass from the environment until the $Q=1$ threshold is reached, upon which the disc starts to `release water' onto the central star. The steady state is that the disc and star mass grow together. %A better analogy is with rainwater (from the `environment') filling a water tank that drains into a swimming pool, where both continue to fill, but such that the ratio of water in the tank and swimming pool remains constant. In Figure~\ref{fig:G1_img1} our tank (the disc) reaches a steady state where it holds 25\% of the mass of water in the swimming pool (the star). The drain between the two ($\dot{M}_\star$) allows most, but not all, of the infalling material to flow onto the star since $\dot{M}_\star = \dot{M}_{\rm inj}/(1 + q)$ from Eq.~\eqref{mdot_star_SG}. The drain shrinks slightly with time as $q$ increases but then admits a constant flux in the steady state when $q$ is constant.

Figure~\ref{fig:comparison} confirms that this simple picture holds in both 1D and 3D simulations. While we have injected mass in an axisymmetric way to minimise the number of free parameters, we expect the same steady state behaviour independent of the way mass is added to the disc. That a steady state is reached means that past work using $\beta$-cooling as a numerical sandbox to understand turbulence, spiral arms, dust aggregation and angular momentum transport driven by gravitational instability can be ported to this new and more realistic framework of infall driven mass-regulated gravitational instability.

\subsection{Observational signatures}
A complication is that infall itself can drive spiral arms in the disc, if the mass is added in a non-axisymmetric way \citep{calcino25}. Hence it is important to be able to distinguish spiral arms driven by infall from those driven by gravitational instability. Infall is a necessary but not sufficient condition for gravitational instability to operate. The most robust way to distinguish the two is to measure the disc mass (e.g. with dynamical disc mass measurements; \citealt{veronesi21,lodato23,martire24,longarini25}) and hence constrain the Toomre~$Q$ parameter directly. %Another possibility is to measure the pattern speed of the spirals \citep[e.g.][]{ren20}, since \citet{calcino25} found that infall can induce spiral arms that are fixed in the observers frame. However since we expect gravitational instability to always be driven by infall, one would expect such spirals to also be present alongside the instability-driven spiral density waves in real discs.

\citet{hall20}, using $\beta-$cooling simulations, showed that gravitational instability can leave a unique signature in channel maps of molecular line emission from discs --- the `GI wiggle'. Essentially spirals produce global scale perturbations in the velocity field which shift emission into neighbouring frequency channels. This characteristic wiggle has been observed by ALMA in Elias 2-27 \citep{paneque21, longarini24} and AB Aur \citep{speedie24} and claimed as evidence for gravitational instability operating in these discs. %However the comparison with numerical simulations based on $\beta$-cooling is problematic because of the mismatch in temperatures discussed in Section~\ref{sec:thermostat}, which argues against gravitational instability \citep{hall18}.
We expect GI-induced wiggles to persist in the infall-driven gravitational instability scenario. However, the amount of angular momentum transported by the spirals --ultimately setting the wiggle amplitude \citep{longarini21,longarini24} -- would no longer be governed by $\beta-$cooling. Instead, it would depend on a combination of the mass injection rate and the disc's aspect ratio. As noted by \citet{longarini24}, if the wiggle amplitude is interpreted as a measure of the angular momentum transport driven by GI -- specifically, \(\alpha_{\rm GI}\) as defined in \citet{longarini24} -- then it remains independent of the mechanism driving self-regulation.

%Infall-driven gravitational instability would be expected to produce similar observational signatures since the low-$m$ modes remain the dominant spiral structure (Section~\ref{sec:morphology}). Hence we expect that a similar large-scale wiggle would be predicted, and would be a more natural interpretation for these discs since infall onto the disc from the environment is also observed \citep{fukugawa04,speedie24}. The similar morphology of the expected spiral arms explains the success of the $\beta$-cooling model in predicting observed signatures, even with unrealistic disc temperatures. However, a detailed prediction of the observational signatures from infall-driven gravitational instability should be performed to check, and one should be careful to distinguish purely infall-driven spirals from infall+gravitational instability driven modes.

\subsection{Setting the initial stages for planet formation}
In this paper we showed that if a protostellar disc undergoes infall driven gravitational instability, it reaches a self-regulated state where star mass, disc mass and accretion rate are related. Large surveys of star-forming regions have revealed robust power-law correlations between stellar mass and disc properties \citep{ansdell16, ansdell17, barenfeld16, alcala17, manara17, manara20, testi22}. These correlation seem to evolve, steepening with time. \cite{somigliana22} showed that these correlations cannot be solely explained by viscous evolution unless an initial correlation is already imprinted in the system’s initial conditions.

Infall-driven gravitational instability is a promising mechanism capable of imprinting correlations in the initial conditions for the viscous evolution of protostellar discs. In particular, we have shown that this mechanism naturally leads to a disc-to-star mass ratio proportional to the disc aspect ratio, while the mass accretion rate scales with  $\alpha$, which itself depends on the stellar mass, aspect ratio, and mass injection rate. To robustly predict the scaling relations of $M_d$ and $\dot{M}_\star$ with stellar mass, one must consider several physical factors that regulate disc structure and evolution during the early phases. For instance, the disc temperature plays a critical role, as it sets the aspect ratio. 

Although in later stages the temperature of a passively irradiated disc is known to depend on the stellar luminosity --- and therefore the stellar mass --- during the earliest phases, accretion luminosity is the dominant heating source \citep{krumholz07,offner09,henebelle22}, effectively setting the thermal structure of the disc. In an infall-driven scenario, this luminosity is tightly linked to the mass injection rate, which in our model is assumed to be constant in time and radially localised. In reality, this is not the case \citep{Cassen81,terebey84,hueso05,Bae13,hartmann18}. Exploring these effects in more detail could provide further insight into the phenomena setting the initial stage of planet formation.

\subsection{Rescaling the injection rate}
Our study is performed with protostellar discs in mind, but our analysis is applicable across a wide mass range, from protostellar discs to discs around black holes where there is irradiation from a central source (which could be the inner disc itself). Since all our simulations are scale-free, they can be used to interpret any accretion disc. The same principle applies to the injection rate, which must be considered in relation to the disc's physical size and the injection radius. The key variable that governs angular momentum transport is not simply $\dot{M}_{\rm inj}$, but rather the dimensionless quantity defined $\dot{\mu}$, as defined in Eq.~\eqref{dotmu}.  

\begin{figure*}
    \centering
    \includegraphics[width=0.475\linewidth]{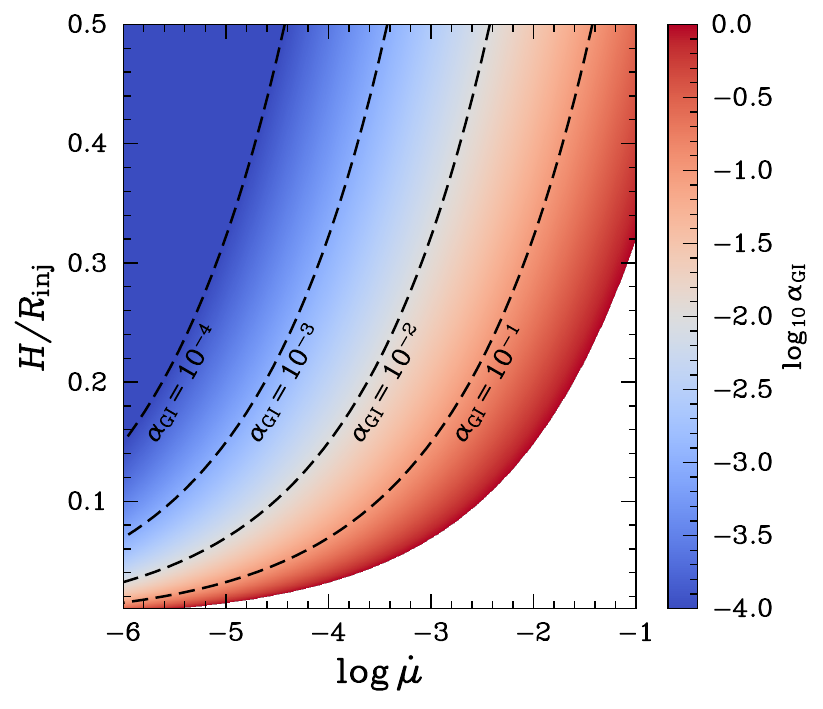}
    \includegraphics[width=0.475\linewidth]{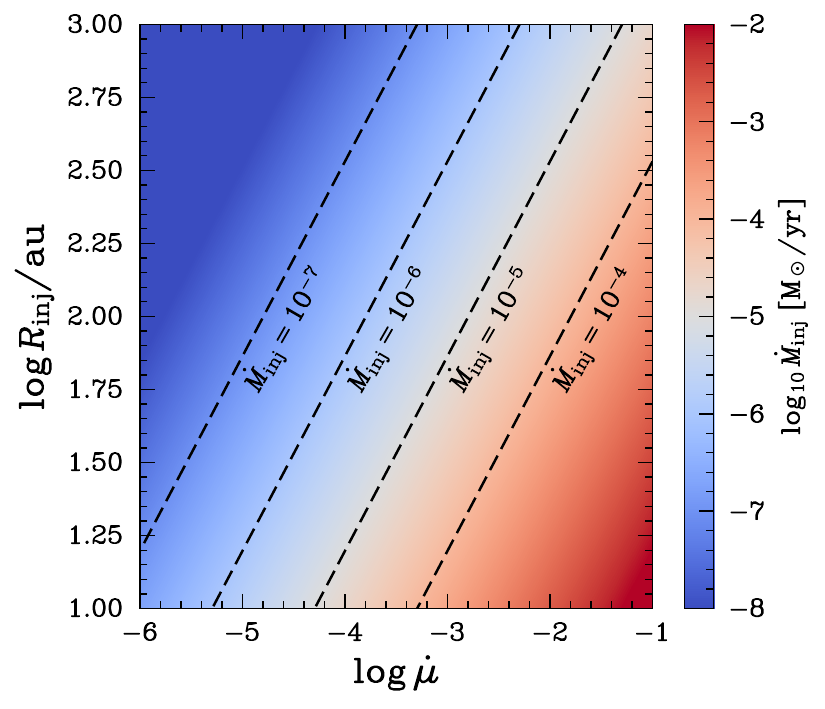}
    \caption{Left panel: $\alpha-$viscosity as a function of $\Gamma$ and the aspect ratio. Right panel: physical injection rate as a function of the injection radius and $\Gamma$.}
    \label{fig:alpha_TH}
\end{figure*}

Figure~\ref{fig:alpha_TH} illustrates how to rescale the \textit{physical} injection radius. The left panel shows the $\alpha$-viscosity as a function of $\dot{\mu}$ and $H/R$, as described in Eq.~\eqref{alpha_TH}, where, for simplicity, the term $(1+q)$ has been omitted. A specific value of $\alpha$ is determined by the combination of $H/R$ and $\dot{\mu}$. To convert $\dot{\mu}$ into physical units --- specifically, an injection rate in solar masses per year — the right panel of Figure~\ref{fig:alpha_TH} provides the corresponding scaling relation based on the injection radius, according to 
\begin{equation}
    \log\dot{M}_{\rm inj}[{{\rm M}_\odot/{\rm yr}}]  \simeq \log \dot{\mu} + \frac{3}{2}\log M_\star[{{\rm M}_{\odot}}]-\frac{3}{2}\log R_{\rm inj}[\rm{au}] + 1.6.
\end{equation}

For instance, in a protoplanetary disc with an aspect ratio of  $H/R = 0.1$  and a viscosity parameter  $\alpha = 10^{-2}$ , the dimensionless mass injection rate is  $\log_{10} \dot{\mu} = -4.5$ . Injecting material at  50  au results in a physical injection rate of  $6 \times 10^{-7} \, M_\odot \, \text{yr}^{-1}$ . A similar calculation applies to AGN discs around black holes. For a disc with  $H/R = 0.05$  and  $\alpha = 10^{-2}$ , the dimensionless injection rate is  $\log_{10} \dot{\mu} = -5.42$ . If mass is injected at  $R_{\rm inj} = 0.1$  pc around a  $10^6 \, M_\odot$  black hole, the corresponding physical injection rate is  $5.5 \times 10^{-2} \, M_\odot \, \text{yr}^{-1}$ .

\section{Conclusions}\label{conclusion}
In this work, we characterised gravitational instability triggered by mass injection through 1D and 3D numerical simulations. Our findings can be summarised as:

\begin{enumerate}
    \item Self-regulation can be achieved in terms of mass rather than temperature, leading to a constant disc-to-star mass ratio over time. We verify that self-regulation occurs in both 1D and 3D simulations and find that the characteristic timescale is the injection time, corresponding to the stellar mass doubling timescale. Additionally, self-regulation is reached more rapidly in 3D than in 1D.  
    \item The 1D evolution code enables us to study the evolution of key global parameters, including \( M_\star \), \( M_d \), their ratio, and the disc radius \( R_d \), as functions of the aspect ratio, injection rate, and injection location. We find that the aspect ratio is the critical parameter, as it sets the maximum mass the disc can sustain.  
    \item 3D SPH simulations show excellent agreement with the 1D code and provide insights into the morphology of the spirals. Unlike the cooling-driven case, the spirals are less flocculent, with an azimuthal power spectrum peaking at the dominant mode and exhibiting little power in higher \( m \) modes. Additionally, we observe a time correlation between different modes, particularly between \( m=2 \) and \( m=3 \).  
    \item Unlike the cooling scenario, the pattern speed of the spiral perturbations in the infall driven GI have a constant pattern speed, matching the angular velocity at the injection location. In this picture, the injection radius is the corotation resonance, and it is possible to identify the inner and outer Lindblad resonances. Spiral waves generate at the injection, and freely propagate inwards since the Toomre parameter is $Q\sim1$, while between the injection and the outer Lindblad resonance the wave is evanescent, since $Q>1$ and $k\in $Im. 
\end{enumerate}

\section*{Acknowledgements}
The authors thank the referee for the constructive report which improved the clarity of the paper. In addition, the authors thank Rebecca Nealon for the de-refinement routine in \textsc{Phantom}, Duncan Forgan for contributing the disc stresses and mode analysis module and Alice Somigliana for helping in the discussion about the initial conditions. The authors finally thank Juan Garrido-Deutelmoser, Andrew Winter, Francesco Zagaria, Jess Speedie and Cat Leedham for useful discussions.

This work has received funding from the European Union’s Horizon 2020 research and innovation programme under the Marie Sklodowska-Curie grant agreement \# 823823 (RISE DUSTBUSTERS project). CL and CJC have been supported by the UK Science and Technology Research Council (STFC) via the consolidated grant ST/W000997/1. GL acknowledges support from PRIN-MUR 20228JPA3A and from the European Union Next Generation EU, CUP:G53D23000870006. KMK acknowledges support NASA under agreement No. 80NSSC21K0593 for the program “Alien Earths.

The simulations presented in this work were performed using the DiRAC Data Intensive service at Leicester (DiAL3), operated by the University of Leicester IT Services, which is part of the STFC DiRAC HPC Facility (www.dirac.ac.uk) within the RAC large project DISCSIM IV and Cambridge Service for Data Driven Discovery (CSD3). 

In this work, we used \textsc{Splash} to create figures of the hydrodynamical simulations \citep{price07} and \textsc{Sarracen} \citep{sarracen} for part of the analysis.

%%%%%%%%%%%%%%%%%%%%%%%%%%%%%%%%%%%%%%%%%%%%%%%%%%
\section*{Data Availability}
The simulations performed in this work will be made available on Zenodo \url{10.5281/zenodo.14906500}.

%%%%%%%%%%%%%%%%%%%% REFERENCES %%%%%%%%%%%%%%%%%%

% The best way to enter references is to use BibTeX:

\bibliographystyle{mnras}
\bibliography{example} % if your bibtex file is called example.bib

% Alternatively you could enter them by hand, like this:
% This method is tedious and prone to error if you have lots of references
%\begin{thebibliography}{99}
%\bibitem[\protect\citeauthoryear{Author}{2012}]{Author2012}
%Author A.~N., 2013, Journal of Improbable Astronomy, 1, 1
%\bibitem[\protect\citeauthoryear{Others}{2013}]{Others2013}
%Others S., 2012, Journal of Interesting Stuff, 17, 198
%\end{thebibliography}

%%%%%%%%%%%%%%%%%%%%%%%%%%%%%%%%%%%%%%%%%%%%%%%%%%

%%%%%%%%%%%%%%%%% APPENDICES %%%%%%%%%%%%%%%%%%%%%

\appendix

\section{1D evolution code}
\label{sec:1Dcode}
We solve the 1D evolution equation for the surface density of a Keplerian disc using a first-order explicit finite-volume update following \citep{bath81}. We work in dimensionless units, where $\tau=t/t_{\rm inj}$, with  $t_{\rm inj}=\text{M}_\odot /\dot{M}_{\rm inj}$, $\sigma = \Sigma / \Sigma(R_{\rm inj})$, $X = R/R_{\rm inj}$ and $D = \nu \times(t_{\rm inj}/R_{\rm inj}^2)$. Equation \ref{diff_eq_dimensional} becomes
\begin{equation}
    \frac{\partial \sigma}{\partial \tau} = \frac{3}{X}\frac{\partial}{\partial X}\left[\sqrt{X}\frac{\partial}{\partial X}\left(\sqrt{X}\sigma D\right)   \right] + \gamma_{\rm inj},
\end{equation}
where $\gamma_{\rm inj} = \dot{\Sigma}_{\rm inj} \times (t_{\rm inj}/\Sigma(R_{\rm inj}))$. We solve the previous equation using the following discretisation 
\begin{equation}
    {\sigma_j^{n+1}} = \sigma_j^n + \frac{3 \Delta t}{X_j}\frac{\sqrt{X_{j+1/2}}  \left( A_{j+1}^n - A_j^n \right) - \sqrt{X_{j-1/2}}  \left( A_{j}^n - A_{j-1}^n \right)}{(\Delta X)^2} + {\gamma}_j \Delta t,
\end{equation}
where 
\begin{equation}
    A_j^n=\sqrt{X_j}\sigma_j^nD_j.
\end{equation}

The code used to evolve the surface density of the disc is publicly available on \url{https://github.com/crislong/discfusion}.

\section{Viscosity prescription for gravitationally unstable discs}\label{app_viscosity}
\begin{figure}
    \centering
    \includegraphics[width=0.85\linewidth]{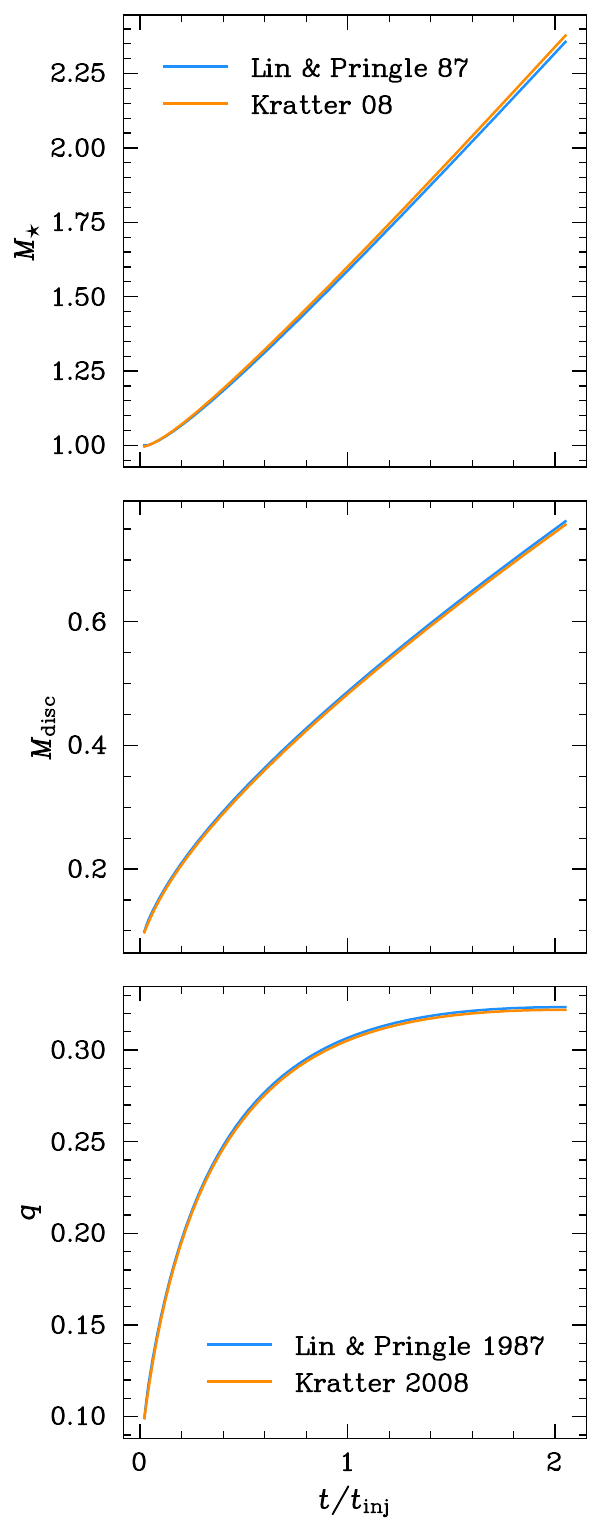}
    \caption{Evolution of star mass, disc mass and disc to star mass ratio in the 1D evolution code with two different viscosity prescriptions, namely \citet{linpringle87} and \citet{kratter08}.}
    \label{fig:viscosity_kratter}
\end{figure}

In this appendix, we assess the robustness of our 1D results against different viscosity prescriptions. We initially adopted the GI viscosity model from \cite{linpringle87}, where angular momentum transport occurs only when $Q<1$. This model is very simple and features an abrupt transition between active and inactive viscosity. To test the sensitivity of our results, we also consider an alternative viscosity prescription from \cite{kratter08}, given by
\begin{equation}
    \alpha_{\rm Kratter} = \max\left\{ 0.14 \left[\left(\frac{1.3}{\max(Q,1)}\right)^2-1\right], 0 \right\}.
\end{equation}
According to this prescription, the transition between active and inactive is shallower. Figure~\ref{fig:viscosity_kratter} shows the evolution of star mass, disc mass and disc to star mass ratio, comparing the two viscosity prescriptions. We find an excellent agreement, meaning that our results are robust.

%\section{Reynolds and gravitational stresses}
%\begin{equation}
%    \mathbf{v}^i=\mathbf{v}_{\rm avg}^i + \delta\mathbf{v}^i
%\end{equation}
%\begin{equation}
%    \mathbf{v}_{\rm avg}^i = \sum_j^{N_{\rm neigh}}\frac{m_j}{\rho_j}\mathbf{v}^j\text{W}(r_{ij},h_j)
%\end{equation}
%\begin{equation}
%    T_{R\phi}^{\rm Re} = - \Sigma \langle\delta v_R\delta v_\phi\rangle
%\end{equation}
%\begin{equation}
 %   \langle\delta v_R\delta v_\phi\rangle^i = \sum_j^{N_{\rm neigh}}\frac{m_j}{\rho_j}\left(\delta v_R\delta v_\phi\right)^j\text{W}(r_{ij},h_j)
%\end{equation}

\section{Spiral tracking}\label{app:tracking}
In this appendix we show the results of the tracking procedure for the simulation \textbf{S3D\_1} and \textbf{S3D\_2}. 

\begin{figure*}
    \centering
    \includegraphics[width=0.495\linewidth]{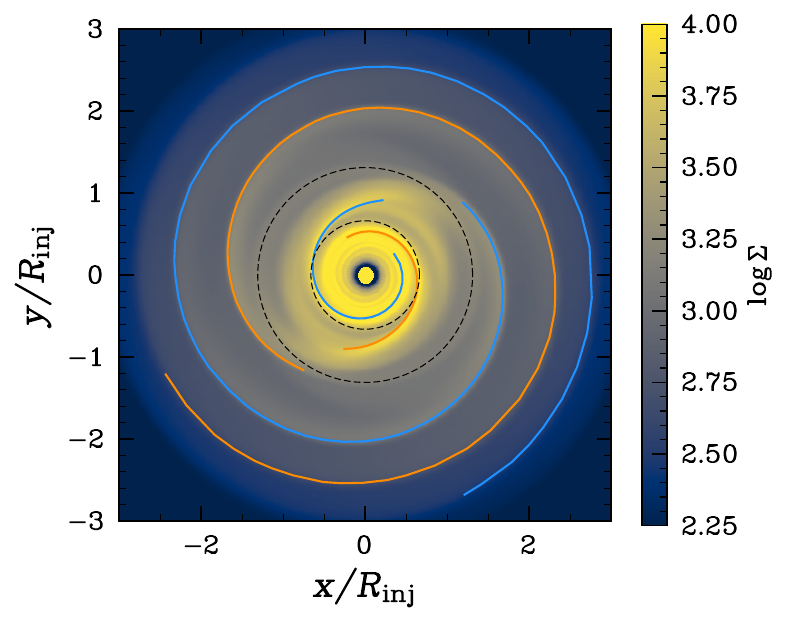}
    \includegraphics[width=0.425\linewidth]{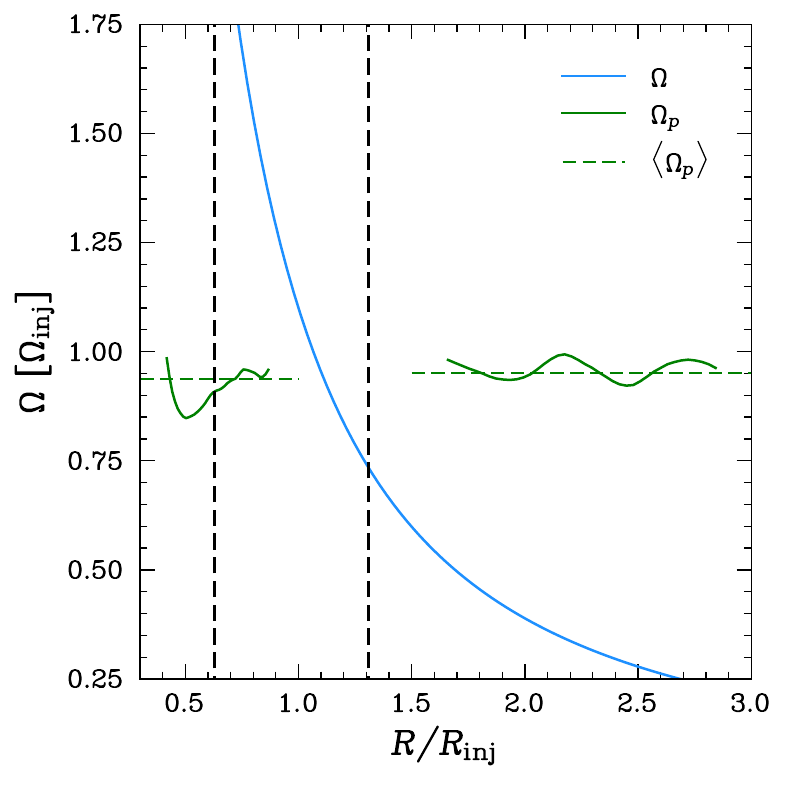}

    \includegraphics[width=0.475\linewidth]{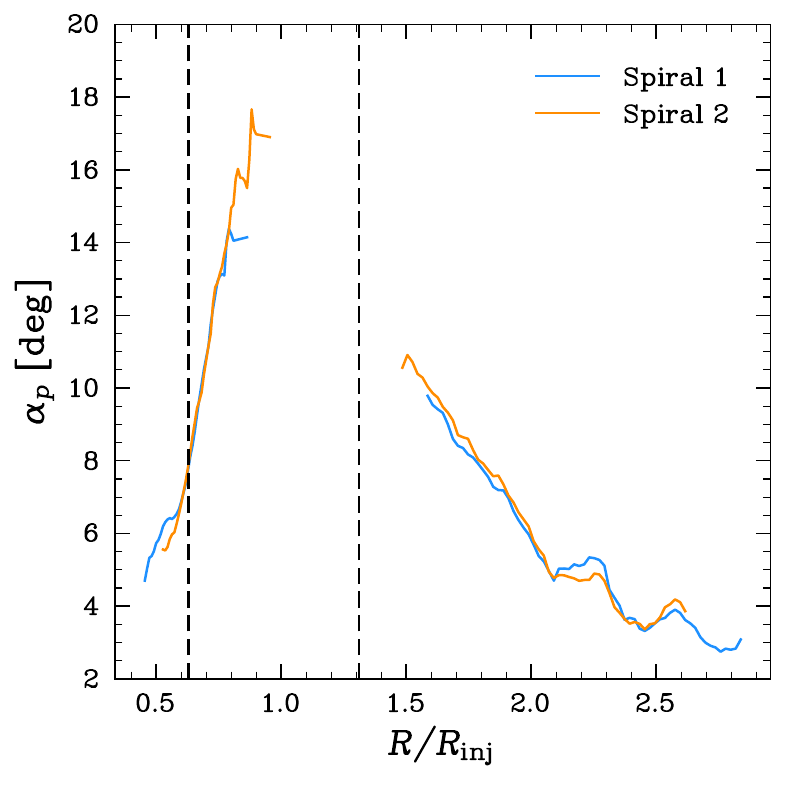}
    \includegraphics[width=0.475\linewidth]{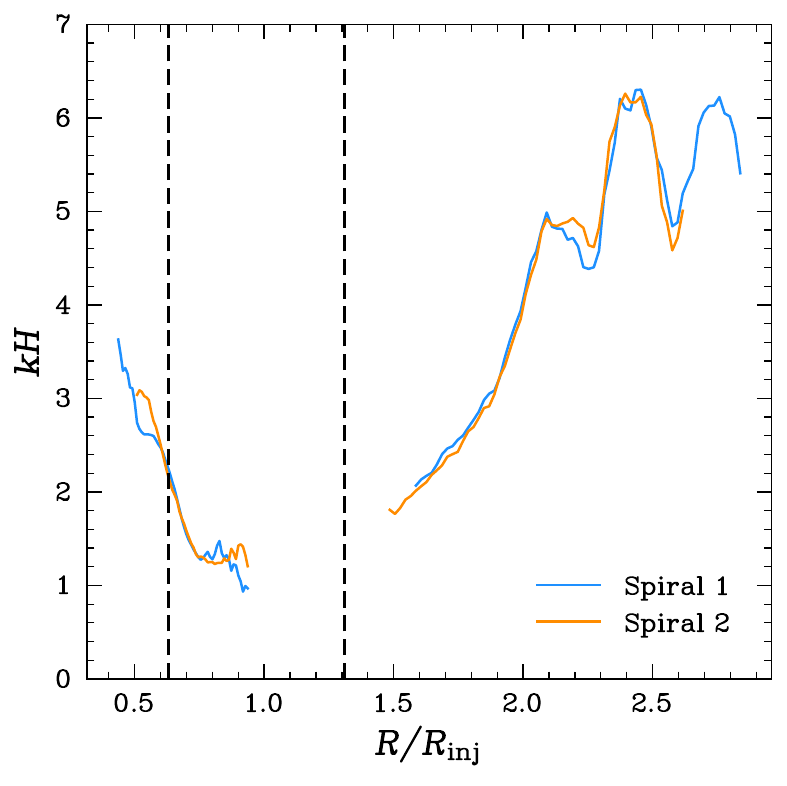}
    \caption{Results of the spiral tracking procedure for the \textbf{S3D\_1} simulation.}
    \label{fig:enter-label}
\end{figure*}

\begin{figure*}
    \centering
    \includegraphics[width=0.495\linewidth]{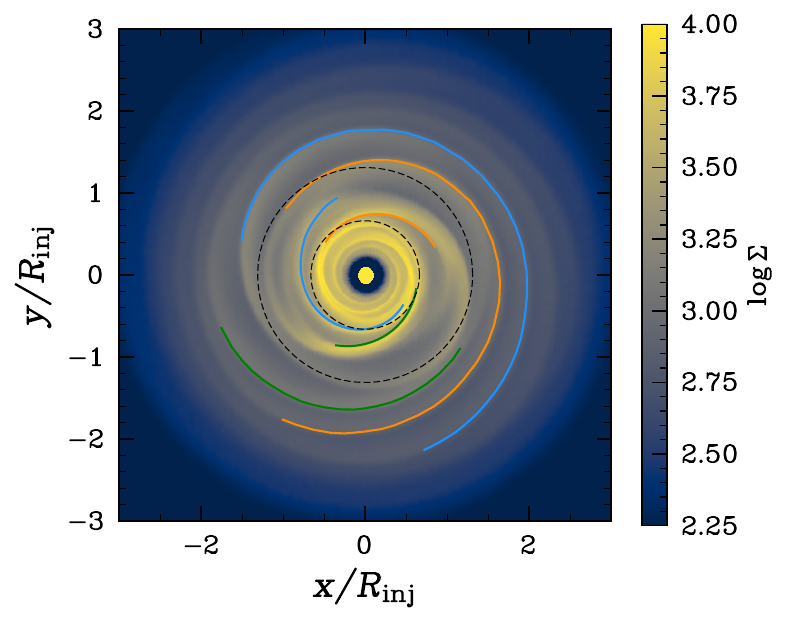}
    \includegraphics[width=0.425\linewidth]{images/pspeedhr01.pdf}

    \includegraphics[width=0.475\linewidth]{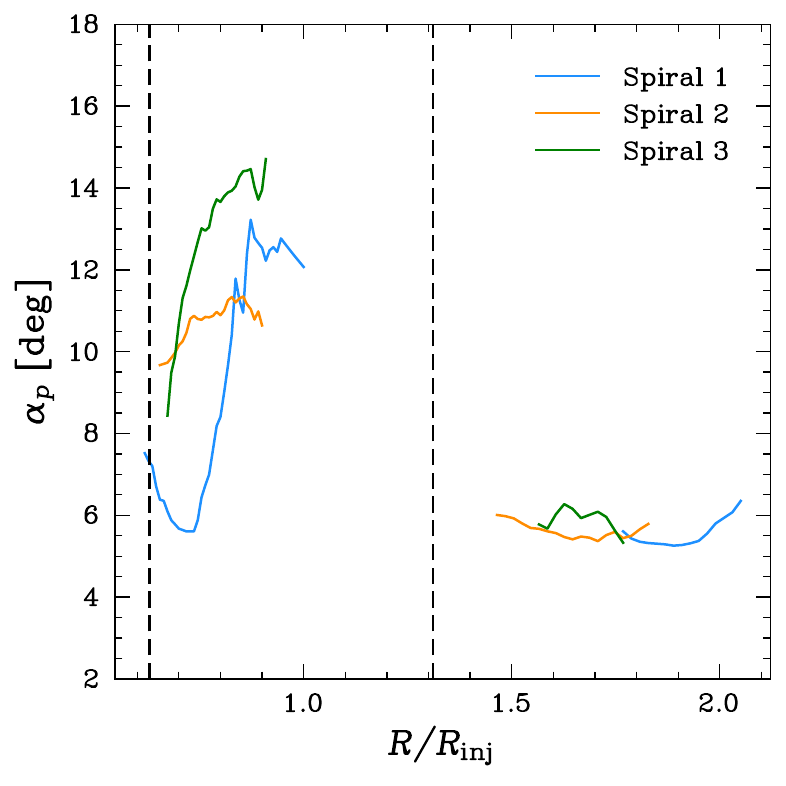}
    \includegraphics[width=0.475\linewidth]{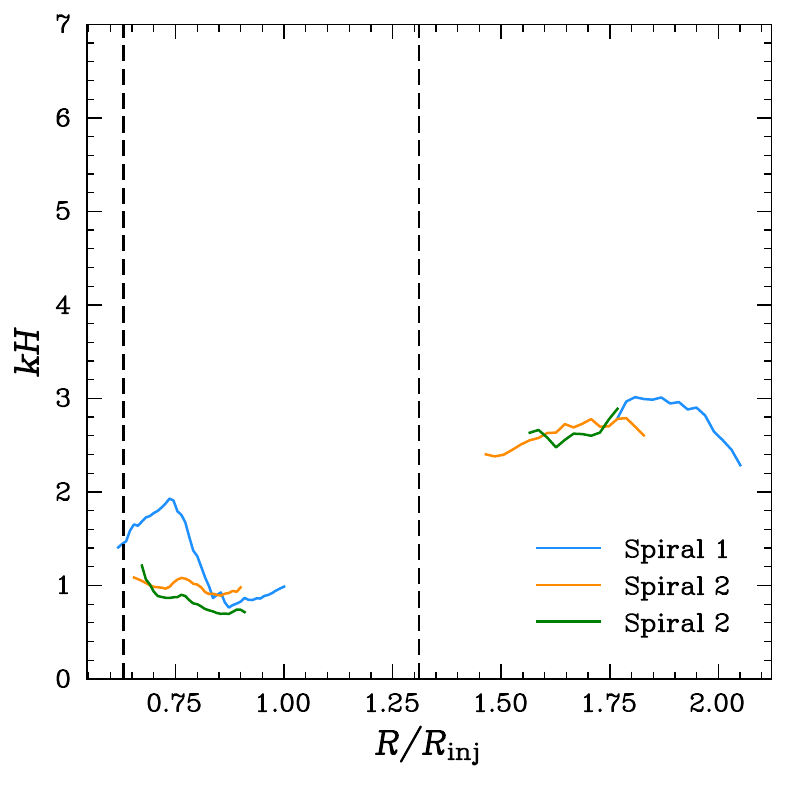}
    \caption{Results of the spiral tracking procedure for the \textbf{S3D\_2} simulation.}
    \label{fig:enter-label}
\end{figure*}

% Don't change these lines
\bsp	% typesetting comment
\label{lastpage}
\end{document}